\documentclass{jfm}
\usepackage{graphicx}
\usepackage{epstopdf, epsfig}
\usepackage{amsmath}
\usepackage[mathscr]{euscript}
\usepackage{ulem}

\usepackage{xcolor,cancel}
\usepackage{caption}
\usepackage{subcaption}

\numberwithin{equation}{section}
\renewcommand{\theequation}{\thesection.\arabic{equation}}
\shorttitle{Stability of dusty flow}
\shortauthor{Anup Kumar, Rama Govindarajan}

\title{Mechanism of instability in non-uniform dusty channel flow}

\author{Anup Kumar
  \corresp{\email{anup.kumar@icts.res.in}},
 \and Rama Govindarajan}

\affiliation{ International Centre for Theoretical Sciences,
Bangalore-560089, India
}

\begin{document}

\maketitle

\begin{abstract}
Particles in pressure-driven channel flow are often inhomogeneously distributed. Two modes of low-Reynolds number instability, absent in Poiseuille flow of clean fluid, are created by inhomogeneous particle loading, and their mechanism is worked out here. Two distinct classes of behaviour are seen: when the critical layer of the dominant perturbation overlaps with variations in particle concentration, the new instabilities arise, which we term overlap modes. But when the layers are distinct, only the traditional Tollmien-Schlichting mode of instability occurs. We derive the dominant critical layer balance equations in this flow along the lines done classically for clean fluid. These reveal how concentration variations within the critical layer cause two the particle-driven instabilities. As a result of these variations, disturbance kinetic energy production is qualitatively and majorly altered. Surprisingly the two overlap modes, though completely different in the symmetry of the eigenstructure and regime of exponential growth, show practically identical energy budgets, highlighting the relevance of variations within the critical layer. The wall layer is shown to be unimportant. We derive a minimal composite theory comprising all terms in the complete equation which are dominant somewhere in the flow, and show that it contains the essential physics. 

When particles are infinitely dense relative to the fluid, the volume fraction is negligible. But for finite density ratios, the volume fraction of particles causes a profile of effective viscosity. This is shown to be uniformly stabilizing in the present flow. Gravity is neglected here, and will be important to study in future. So will transient growth of perturbations due to non-normality of the stability operator, in a quest for the mechanism of transition to turbulence.
\end{abstract}


\section{Introduction}

The dynamics of fluids laden with suspended particles has been a subject of investigation for decades. When the particles are extremely small and in great number, the term ``dusty flow" is appropriate. Dusty shear flows are ubiquitous: occurring in environmental phenomena like dust storms, snow avalanches and sediment transport in rivers, and in industrial processes like the manufacture of fertilizers and various powders. Whether such a flow will be laminar, turbulent or in an unsteady transitional state is of great interest for a variety of reasons, and the first step is to study the stability of the laminar base state. 

\cite{saffman1962stability} was the first to propose a formulation for the stability of a pressure-driven laminar channel flow, of a fluid containing dust particles in dilute suspension. The dust particles were uniformly distributed across the channel width. Subsequently, \cite{michael1964stability} conducted numerical computations, validating the conclusions of \cite{saffman1962stability}. \cite{isakov1995stability} extended the study of \cite{michael1964stability} with improved numerical accuracy. \cite{boronin2008stability} studied uniform particle loading with a finite volume fraction modeled by a corrected Stokes drag including the effects of viscosity variations due to perturbations in particle concentration, and found destabilization compared to the dusty-gas results of \cite{isakov1995stability}. 
In a later study, \cite{klinkenberg2011modal} noted that the critical Reynolds number increases to high levels with strengthening of loading in a uniform particle distribution. Nevertheless, at a Reynolds number of a few thousand, loading of particles can enhance the transient growth for three-dimensional perturbations. \cite{nath2024instability} found that in simple shear flow, non-uniformly distributed particles destabilize the flow through an inviscid mechanism. This is in contrast to our system of plane channel flow, where we show that destabilization is by a viscous mechanism.

Small particles suspended in channel or pipe flow normally do not occur with uniform probability everywhere \citep{MATAS_MORRIS_GUAZZELLI_2004}. They tend to concentrate in certain relatively thin regions of the flow. The location of concentration depends on different flow and loading conditions, and examples are available in the experiments of  \cite{Snook_Butler_Guazzelli_2016}. The early experiments of \cite{segre1961radial,segre1962behaviour} showed that particles, when homogeneously distributed in a pipe, undergo inertial cross-stream migration, caused by lift forces and the wall, and tend to accumulate within an annular region, located at a certain radial distance. \cite{Saffman_1965} calculated the lift force for a small solid particle in unbounded linear shear.  
 \cite{cox1968lateral} included considerations of the wall, and of shear variation. The review of \cite{cox1971suspended} provides the equilibrium radial variation of particle concentration in a range of conditions in a pipe. For two-dimensional channel flow, studied here, \cite{ho1974inertial} offered the first theoretical explanation for non-uniform particle loading, due to the wall-induced lift force and the shear-gradient lift force. For neutrally buoyant particles, they found two equilibrium points: an unstable point at the channel centerline and stable points located $\pm 0.6$ times the half-channel width from the center. Their calculations were for creeping flow in the channel, namely for channel Reynolds number $R \ll 1$, as well as the particle Reynolds number $Re_p$ being significantly smaller than $R$. \cite{schonberg1989inertial,asmolov1999inertial} revealed a wallward shift of the stable equilibrium points for increasing $R$. For particles of a finite size relative to the channel width, \cite{anand2023inertial} found an additional equilibrium point closer to the centerline.

Thus an inhomogeneous equilibrium particle distribution with a relatively thin particle-containing layer is natural in channel flow, though its location depends on several factors. The question is whether this arrangement remains stable to an accumulation of particles, or whether such accumulation, when sufficiently high, can cause the laminar flow to undergo instability.
We adopt a gaussian particle distribution profile to model experimental observations and theoretical findings.
Following the findings of \cite{klinkenberg2011modal} and the calculations of many, we may take the lift force at the equilibrium location, on a sufficiently small particle, to be negligibly small compared to the Stokes drag.

\cite{rudyak1996stability,rudyak1997hydrodynamic} investigated the effects of inhomogeneous gaussian particle loading and found low Reynolds number instabilities of the kind we discuss here. A channel loaded with particles where particle concentration tapers off towards the walls was shown by \cite{boronin2009hydrodynamic} to support instability at zero Reynolds number. Incidentally we found the instabilities of \cite{rudyak1996stability,rudyak1997hydrodynamic} independently, since we only learned about that work recently. They noticed that the critical layer lies close to the particle-laden layer in these instabilities, but did not provide the mechanism which generates the new instabilities. The mechanism is the subject of the present paper. Along the lines of the famous classical derivation of \cite{lin1945stabilitya,lin1945stabilityb,lin1946stability_part3} for a clean fluid, we derive the critical-layer and wall-layer equations for dusty parallel shear flow. The critical layer equations make it obvious how the inhomogeneity of particle loading enters the leading-order physics. We derive a minimal composite equation containing all the leading-order terms and show that it contains the essential overlap physics. Our energy budget study and the eigenfunctions support our findings, and directly show how production of perturbation kinetic energy is altered in the critical layer. Our study demarcates two distinct classes of stability behaviour: one where the critical layer overlaps the layer where the particle concentration is non-constant, and another where the two layers lie away from each other. Two modes of overlap instability occur in the former.

Incidentally the earlier study on inhomogeneous particle loading only briefly mentions the numerical method, but provides no details of the discretisation, or the level of accuracy of the solutions. In order to achieve reasonable accuracy, we find that a high grid-resolution is needed within the particle-laden layer. 
 
Introducing particles into the flow exacerbates the complexity of the transition to turbulence \citep{mueller2010rheology}. \cite{Guazzelli2003_p1,Guazzelli2003_p2}, in pipe flow experiments, observed that adding particles at a significant volume fraction can delay or advance the transition to higher or lower Reynolds number, based on whether the particles are extremely small or somewhat larger, with a minimum in the transition Reynolds number being attained at a particular volume fraction. Numerical and experimental studies conducted by
 \citet{Guazzelli2003_p1, yu2013numerical,lashgari2014laminar,wen2017experimental} demonstrate that transition to turbulence occurs smoothly, with velocity and pressure fluctuations increasing gradually. This suggests that particles can alter the nature of the transition and the resulting turbulence state.

 Whether or not the transition occurs due to exponentially growing modes, the first step in understanding the transition to turbulence is to understand the mechanism causing linear stable and unstable eigenmodes to exist. We conduct this study below.

\section{The governing equations and their solution}
\begin{figure}
  \centerline{\includegraphics[width=0.95\textwidth]{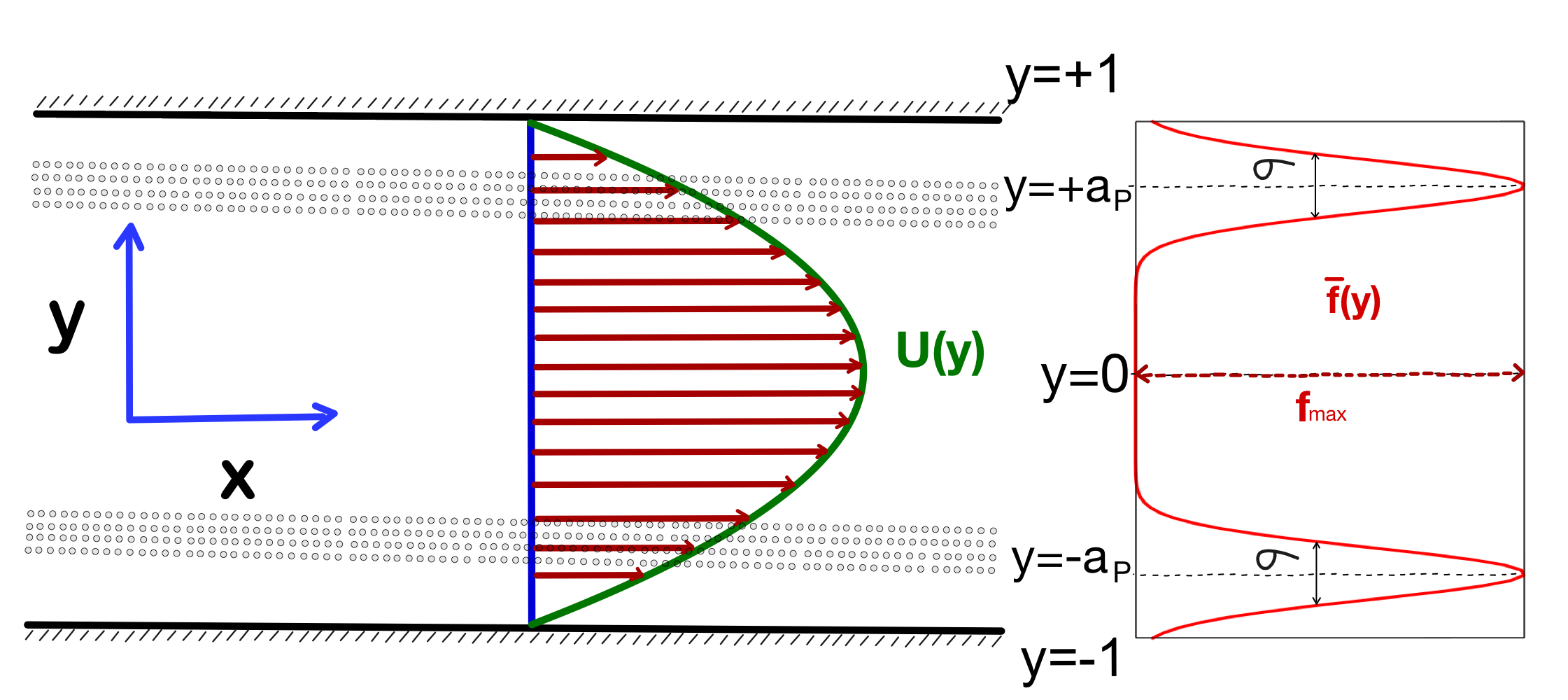}}
  \caption{Schematic of the flow under consideration. The walls are situated at $y=\pm 1$, with the green curve and red arrows representing the mean velocity profile, $U(y) = 1 - y^{2}$. The particles are concentrated around $y = \pm a_p$ within a band of size $\sigma$. The mean particle mass fraction, $\bar f$, given by equation (\ref{feqn}) is depicted on the right. Note that the volume fraction that heavy particles occupy will be much less.}
\label{fig:schematic}
\end{figure}
\subsection{Description of the system}
We investigate here a dilute suspension of particles in a pressure-driven channel flow, a schematic of which is shown in figure \ref{fig:schematic}. The impact of this suspension on the flow is characterized by a two-way coupling, modeled using the formulation of \cite{saffman1962stability}, specifically through the application of Stokes drag, with the addition of viscosity variations due to particle concentration. The viscosity variation terms are derived from \cite{govindarajan2004effect}. The particulate suspension is treated as a continuous medium whose dynamics is describable by a field equation. The momentum balance and continuity equations for the fluid respectively are  
\begin{equation}
\rho_{f}\left(\frac{\partial \mathbf{u_d}}{\partial t_d}+ \mathbf{u_d} \cdot \mathbf{\nabla_d}\mathbf{u_d}\right)=-\mathbf{\nabla_d} p_d + \mathbf{\nabla_d}\left[\mu^{tot}_d\cdot\left(\mathbf{\nabla_{d}}\mathbf{u_{d}}+\left( \mathbf{\nabla_{d}}\mathbf{u_{d}} \right)^T\right)\right]
+K N\left(\mathbf{v_d}-\mathbf{u_d}\right),
    \label{NS_eq}
\end{equation}
\begin{equation}
     \mathbf{\nabla_d}\cdot\mathbf{u_d}=0,
     \label{Continuity_eq}
\end{equation}
while the particle suspension satisfies momentum balance and continuity respectively given by
\begin{equation}
    m N\left(\frac{\partial \mathbf{v_d}}{\partial t_d}+ \mathbf{v_d} \cdot \mathbf{\nabla_d}\mathbf{v_d}\right)=-KN\left(\mathbf{v_d}-\mathbf{u_d}\right),  
    \label{particle_eq}
\end{equation}
\begin{equation}
    \frac{\partial N}{\partial t_d}+\mathbf{\nabla_d}\cdot\left(N \mathbf{v_d}\right)=0.
\label{particle_continuity_eq}
\end{equation}
Here, the subscript $d$ represents a dimensional variable and $\rho_{f}$ is the dimensional density of the fluid. The total dimensional viscosity $\mu^{tot}_d=\mu_f+\mu_p$, where $\mu_f$ is the dimensional viscosity of the fluid and $\mu_p$ is the contribution to viscosity due to the particles. $m$ and $\tau=m/K$ are the mass and relaxation time of a spherical dust particle, $N$ is their number density per unit volume. The quantity $K$ is the drag coefficient given by $6\pi r\mu_f$ for a sphere of radius $r$.  We establish the $x$-axis to be aligned with the channel centerline, the $y$-axis to be oriented in the wall-normal direction, and the $z$-axis to be oriented perpendicular to the plane of the figure. The fluid velocity is given by $\mathbf{u}=(u_x,u_y,u_z)$. The particles are assumed to be continuously distributed in the flow, and to have a continuous velocity variation in space and time, and their dynamics may thus be described as a field, with $\mathbf{v}=(v_x,v_y,v_z)$. Our analysis thus precludes situations of diverging number density and of particle collisions. The mass fraction of particles in the suspension is
\begin{equation}
    f=mN/\rho_{f}=\frac{4\pi Nr^3}{3}\frac{\rho_p}{\rho_f}.
    \label{mass_frac_ex}
\end{equation} 
The density of the solid making up the particles is $\rho_p$. Unless otherwise specified, we work in the limit of $\rho_p/\rho_f \to \infty$, so the volume fraction occupied the the particles is negligible. 
We non-dimensionalize equations (\ref{NS_eq}--\ref{particle_continuity_eq}) using the channel-centerline mean velocity, $U_m$, the half-channel width, $H$, and the viscosity $\mu_f$ of the fluid as scales.
In dimensionless coordinates, the channel walls are positioned at $y=\pm 1$, with the suspended particles concentrated in around $y=\pm a_p$.
We prescribe a mean dust mass-fraction profile 
\begin{equation}
\bar{f}(y)=f_{max}\left[\exp\left\{\frac{-(y-a_p)^2}{2\sigma^2}\right\}+\exp\left\{\frac{-(y+a_p)^2}{2\sigma^2}\right\}\right],
\label{feqn}
\end{equation}
with particles concentrated in two layers of thickness $\sigma$, but our numerical method is general and suitable for any desired particle concentration profile. The location $a_p$ of the maximum in particle concentration is an important parameter. If the same particles were to be uniformly distributed across the channel, the loading would be given by
\begin{equation}
    f_{ave}=\frac{\int_{-1}^{+1}\bar{f}(y;\sigma,a_p)dy}{\int_{-1}^{+1}dy} \simeq \sqrt{2\pi}f_{max} \sigma.
\label{f_eff}
\end{equation}
The Reynolds and Stokes numbers, which will emerge out of the non-dimensionalisation, are given respectively by    
\begin{equation}
        R \equiv \frac{HU_m}{\mu_f/\rho_{f}} \qquad {\rm and} \qquad S \equiv \frac{\tau}{\rho_{f} H^{2}/\mu_f} = \frac{2}{9}\frac{r^2}{H^2}\frac{\rho_p}{\rho_f}.
        \label{nondim}
\end{equation}
 In terms of the average density in the flow, we may also define an effective Reynolds number $R_{eff}=(1+\bar{f}_{ave})R$.
These two quantities, the Reynolds number R and the Stokes number S, along with the thickness $\sigma$ of the particle-laden layer, the mass loading for a given $\sigma$ as measured by $f_{max}$, and the location $a_p$ of the maximum in particle concentration are the parameters which determine this problem.

\subsection{Linearized equations}
 
After non-dimensionalising, we split all quantities in equations (\ref{NS_eq}--\ref{particle_continuity_eq}) into their basic and fluctuating parts, as
$\mathbf{u} = \mathbf{U}+\mathbf{\hat{u}}$, $\ \mathbf{v} = \mathbf{U}+\mathbf{\hat{v}}$, $ \ p = P+\hat{p}$, $ \  f=\bar{f}+\hat{f}$ and $\mu^{tot}=\bar\mu + \hat \mu$. 
Here a hat represents a perturbation quantity, while an upper case or overbar denotes a mean quantity. In parallel shear flows, we have $\mathbf{U}=U(y)\mathbf{e_x}$, where $\mathbf{e_x}$ is a unit vector in the streamwise direction, and $\bar f=\bar f(y)$. For small particulate volume fraction, the local viscosity is linearly related to the local particle concentration, as
\begin{equation}
    \mu^{tot}_d=\mu_f\left[1+\dfrac{f}{\gamma}\right],
    \label{einstein}
\end{equation} 
where $\gamma \propto \rho_{p}/\rho_{f}$. In accordance with Einstein's law, we take the proportionality constant to be $0.4$. In the limit of infinite $\gamma$, the dimensional viscosity remains at $\mu_f$ everywhere.

For a clean parallel shear flow without particles, \cite{squire1933stability} had shown that for every three-dimensional perturbation mode satisfying the stability equations, there exists a corresponding two-dimensional perturbation mode at a lower Reynolds number, displaying the same growth rate. \cite{saffman1962stability} had shown that Squire's theorem applies in the case of a dusty channel with uniform particle loading. In Appendix A we show that Squire's theorem may be extended to apply for non-uniform particle loading, including viscosity variations. Therefore, while a nonmodal study would require us to study three-dimensional perturbations, since our purpose is to obtain linear instability at low Reynolds number, it is sufficient to perform a two-dimensional calculation. In fact results for any given three-dimensional single mode may be obtained directly from an equivalent two-dimensional one by simple rescaling.

The perturbation quantities are written as integrals of normal modes, with each mode given by:
\begin{equation}
(\mathbf{\hat{u}, \hat{v}}, \hat{f},\hat{\mu})=\frac{1}{2}\left[(\mathbf{u}(y),\mathbf{v}(y),f(y),\mu(y))\exp\left\{i\alpha(x-ct)\right\}+{\rm c.c.} \right].
\label{normal_mode}
\end{equation}
The two-dimensional equations for linear perturbations, after appropriate elimination and reduction (see detailed derivation in Appendix A), can be written in terms of the perturbation streamfunction $\psi(y)$ and the perturbation viscosity $\mu(y)$ as:
\begin{equation}
\begin{aligned}
    \bigg[(U_{*} - c)(D^2 - \alpha^2) &- U_{*}''\bigg]\psi + D(J f' \psi) 
    = \frac{1}{i\alpha R}\bigg[\bar{\mu}(D^2 - \alpha^2)^2 + 2\bar{\mu}' D^3 + \bar{\mu}'' D^2 \\
    &- 2\alpha^2 \bar{\mu}^{\prime} D + \alpha^2 \bar{\mu}''\bigg]\psi + \frac{1}{R}\left[U' D^2 + 2U'' D + U''' + \alpha^2 U'\right]\mu,
\end{aligned}
\label{eq:visc_strat}    
\end{equation}
and 
\begin{equation}
    -(U-c)\gamma\mu+\bigg[-i\alpha RS{\cal M}^{2}U^{\prime}\bar{f}^{\prime}+ ({\cal M}\bar{f})^{\prime}\bigg]\psi=0,
    \label{eq:visc_part_conti}
\end{equation}
where 
\begin{equation}
    U_{*}\equiv U+J\bar{f}, \quad {\cal M}= \frac{1}{1+i\alpha (U-c)SR},\quad J=(U-c){\cal M},
\end{equation}
\begin{equation}
\begin{aligned}
    & u_y=-i\alpha\psi,\quad u_{x}=D\psi, \quad v_{x}={\cal M} u_{x}-({\cal M}^{2}SRU^{\prime})u_{y},\quad v_y={\cal M} u_y.
    \label{visc_para_definitions}
\end{aligned}
\end{equation}
The operator $D$ is defined as $D = d/dy$, and a prime denotes a derivative in $y$ of a mean quantity. The $v_x$ and $v_y$ in equation (\ref{eq:visc_part_conti}) can be expressed in terms of $\psi$ using equation (\ref{visc_para_definitions}). The boundary conditions are: 
\begin{equation}
   \psi(y=\pm 1)=D\psi(y=\pm 1)=\mu(y=\pm1)=0
   \label{boundary_conditions}
\end{equation}
For given mean flow $U(y)$ streamwise wavenumber $\alpha$, base particle loading $\bar f(y)$, particle to fluid density ratio, and fixed Reynolds and Stokes numbers, equations \ref{eq:visc_strat} to \ref{boundary_conditions} define an eigenvalue problem, which yields a spectrum of eigenvalues $c$ and corresponding eigenfunctions, $(\psi(y),\mu(y))$. If even one eigenvalue has a positive imaginary part, i.e., $c_{im}> 0$, we have an exponential growing mode.

In the limit of $\gamma \to \infty$, we have $\bar\mu=1$ and $\mu=0$, so equation (\ref{eq:visc_strat}) becomes
\begin{equation}
      \left[(U_{*}-c)(D^2-\alpha^2)-U_{*}^{\prime\prime}\right]\psi+(J f^{\prime})^{\prime}\psi+(Jf^{\prime})D\psi=\frac{1}{i\alpha R}(D^2-\alpha^2)^2 \psi
      \label{generalised_ev_prob}
\end{equation}
and is now decoupled from equation (\ref{eq:visc_part_conti}). 
When there is no particulate suspension, we have $\bar{f} = 0$, and the system (\ref{generalised_ev_prob})  reduces to the well-known Orr-Sommerfeld equation
\begin{equation}
\left[U(D^2 - \alpha^2) - U''  + \frac{i}{\alpha R}  (D^2 - \alpha^2)^2  \right] \psi = c \left(D^{2}-\alpha^{2}\right)\psi.
\label{os}
\end{equation}
In the case of a homogeneous suspension ($\bar{f} = \text{constant}$) along with $\gamma \to \infty$, the system reduces to that of \cite{saffman1962stability}. 

\subsection{Balance of perturbation kinetic energy}

Whenever the flow is unstable, there is an exponential increase in perturbation kinetic energy. It is useful to derive the positive and negative contributors to this quantity. To do this, we multiply the linear equations for the fluid flow (in $\mathbf{\hat{u}}$) and for the particulate flow (in $\mathbf{\hat{v}}$), given as equations (\ref{NS_eq_linear}) and (\ref{particle_eq_linear}) in Appendix 1, by the respective complex conjugates $\mathbf{\hat{u}^{*}}$ and $\mathbf{\hat{v}^{*}}$. Upon averaging over a wavelength in the streamwise direction, we derive the evolution of perturbation kinetic energy $\hat{E}$ to be described by
\begin{equation}
\begin{aligned}
\partial_t \int \hat{E} \, dV = & - \int \frac{\partial U_i}{\partial x_j} \hat{u}_i \hat{u}_j^ , dV -\frac{1}{R} \int \bar{\mu} \left|\partial_i \hat{u}_j\right|^2 \, dV \\
& -\int f\frac{\partial U_i}{\partial x_j} \hat{v}_i \hat{v}_j dV  -\frac{1}{S R} \int f \left|\hat{u}_i - \hat{v}_i\right|^2 \, dV\\
& -\dfrac{1}{R}\int\frac{\partial^2 \bar{\mu}}{\partial x_j \partial x_i}   \hat{u}_i \hat{u}_j dV -\dfrac{1}{R}\int\frac{\partial U_{j}}{\partial x_{i}} \hat{\mu} (\partial_j \hat{u}_i+\partial_i \hat{u}_j)dV
\end{aligned}
\label{energy_eq}
\end{equation}
where 
\begin{equation}
\hat{E}=\frac{1}{2}(\hat{u}_{i}^{2}+\bar{f}\hat{v}_{i}^{2}), 
\end{equation}
and $V$ indicates a volume of fluid extending from wall to wall and over one perturbation wavelength in the streamwise direction. We then introduce the normal-mode forms of the perturbations, given by equation (\ref{normal_mode}), into equation (\ref{energy_eq}), and average over the streamwise direction $x$, to get
\begin{equation}
\begin{aligned}
2\alpha c_{im} \int E \, dy = & -\frac{1}{4} \int \frac{\partial U_i}{\partial x_j} \left(u_i u_j^* + u_i^* u_j\right) \, dy -\frac{1}{2R} \int \bar{\mu}\left|\partial_i u_j\right|^2) \, dy \\
& -\frac{1}{4} \int f\frac{\partial U_i}{\partial x_j} \left(v_i v_j^* + v_i^* v_j\right) \, dy  -\frac{1}{2S R} \int f \left|u_i - v_i\right|^2 \, dy \\
& -\frac{1}{4R}\int \frac{\partial^2 \bar{\mu}}{\partial x_j \partial x_i}  \left( u_i^* u_j+ u_i u_j^*\right)dy \\
& -\frac{1}{4R}\int \frac{\partial U_{j}}{\partial x_{i}} \{\mu (\partial_j u_i^*+\partial_i u_j^*)+\mu^{*} (\partial_j u_i+\partial_i u_j)\}dy
\\
& \equiv \int (W_{+}-W_{-}+W_{p+}-W_{p-}+W_{\mu,1}+W_{\mu,2})dy,
\end{aligned}
\label{energy}
\end{equation}
where 
\begin{equation}
E(y)=\frac{1}{4}(u_{i}(y)^{2}+\bar{f}v_{i}(y)^{2}),
\end{equation}
and $W_+(y)$ and $W_-(y)$ respectively are the production and dissipation of perturbation kinetic by the fluid, while $W_{p+}(y)$ and $W_{p-}(y)$ respectively give the production and dissipation of perturbation kinetic energy of the particles. The last two terms $W_{\mu,1}$ and $W_{\mu,2}$ arise due to viscosity stratification.

\subsection{Numerical method}
\begin{figure}
  \centerline{\includegraphics[width=0.85\textwidth]{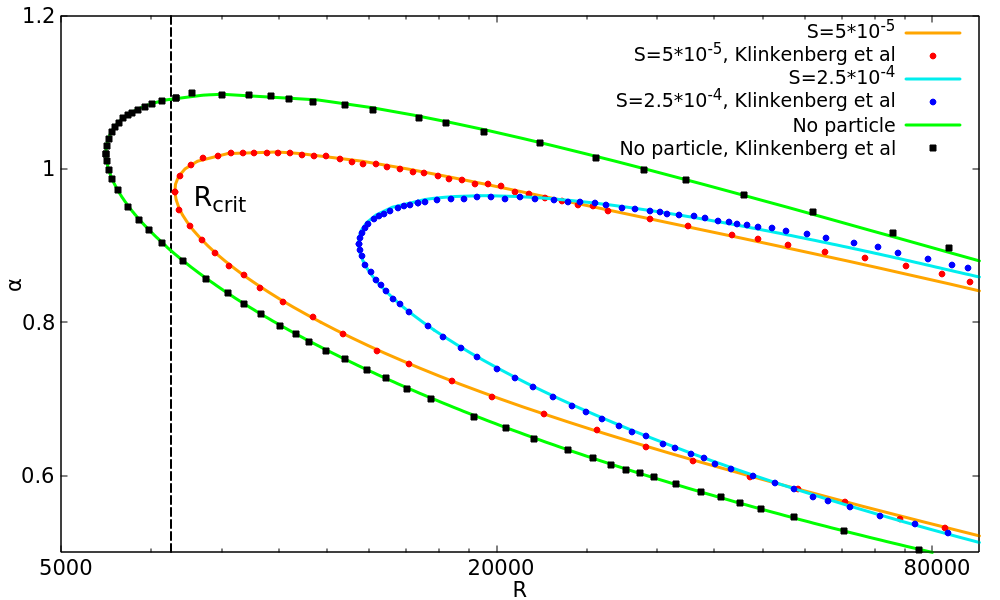}}
  \caption{Validation for the case of uniform particle loading, in the form of neutral stability curves, with $\bar f$ = 0.05, $S=5\times10^{-5}$, and $S=2.5\times10^{-4}$. Symbols correspond to \cite{klinkenberg2011modal}, while solid lines are from present computations. The region within the curves is unstable. The black dotted vertical line marks the minimum Reynolds number for $S=5 \cdot 10^{-5}$ at which instability is seen, termed the critical Reynolds number $R_{crit}$.}
\label{fig:validation_neutral_curves}
\end{figure}
We employ the Chebyshev spectral collocation method to discretize the system given by equations \ref{eq:visc_strat} to \ref{boundary_conditions} at $n$ discrete points in the domain. 
The Chebyshev collocation points, defined as $y_{Cheb, j}=\cos[\pi j/(n-1)],\quad j=0,1,2,3,....,n-1$, are naturally clustered close to the walls. Such a discretisation would resolve the near-wall region well, where  variations are large. But it would leave the particle layer, where too variations are large, not well resolved. In order to get results insensitive to the number of collocation points, we need to employ a stretching function to cluster a sufficient number of grid points into the particle-laden layer. Such a stretching function was used in \cite{govindarajan2004effect} in a different context, and works well in the present situation as well. It is given by
\begin{equation}
    y_j=\frac{a}{\sinh \left(b y_b\right)}\left[\sinh \left\{\left(y_{Cheb,j}-y_b\right) b\right\}+\sinh \left(b y_b\right)\right], 
    \label{stretch_fac}
\end{equation}
where 
\begin{equation}
  y_b=\frac{1}{2b} \log \left[\frac{1+\left(e^b-1\right) a}{1+\left(e^{-b}-1\right) a}\right] \nonumber
\end{equation}
is a constant, $a$ signifies the location around which clustering is desired, and $b$ serves to determine the level of clustering. Once written in discrete form, each boundary condition may be applied by replacing one row of the discrete system appropriately. To solve equation (\ref{generalised_ev_prob}) after discretisation, we utilize the LAPACK FORTRAN package. For our all simulations, we use $n=81$, and verify our answers with $n=121$.

Since the chosen mass fraction profile corresponds to the clustering of particles in the vicinity of $y=\pm a_p$, we select $a$ to be equal to $\pm a_p$ and set $b$ to the value of $2$ or $4$ in equation (\ref{stretch_fac}). We obtain eigenvalues correct to five decimal places for the most part, and at least to four places everywhere. At Stokes number of $10^{-2}$ or higher, however, the accuracy drops to three decimal places, and we do not venture into this regime to make our conclusions.

To validate our approach, we first perform computations using a uniform particle profile across the channel. Figure \ref{fig:validation_neutral_curves} shows neutral stability boundaries provided by \cite{klinkenberg2011modal}, compared to present computations. The agreement is excellent for two different particle Stokes numbers as well as for the clean channel. The mode of instability which appears in all these cases is the traditional Tollmien-Schlichting (hereafter TS) instability, which is modified by the introduction of particles.

We are now in a position to study the instability mechanism. In the following section we derive a minimal equation set which allows us to highlight the basic physics.

\section{A minimal composite theory for particulate shear flow stability}

\begin{figure}
       \centering
        \begin{subfigure}[b]{0.48\textwidth}
         \centering
         \includegraphics[width=\textwidth]{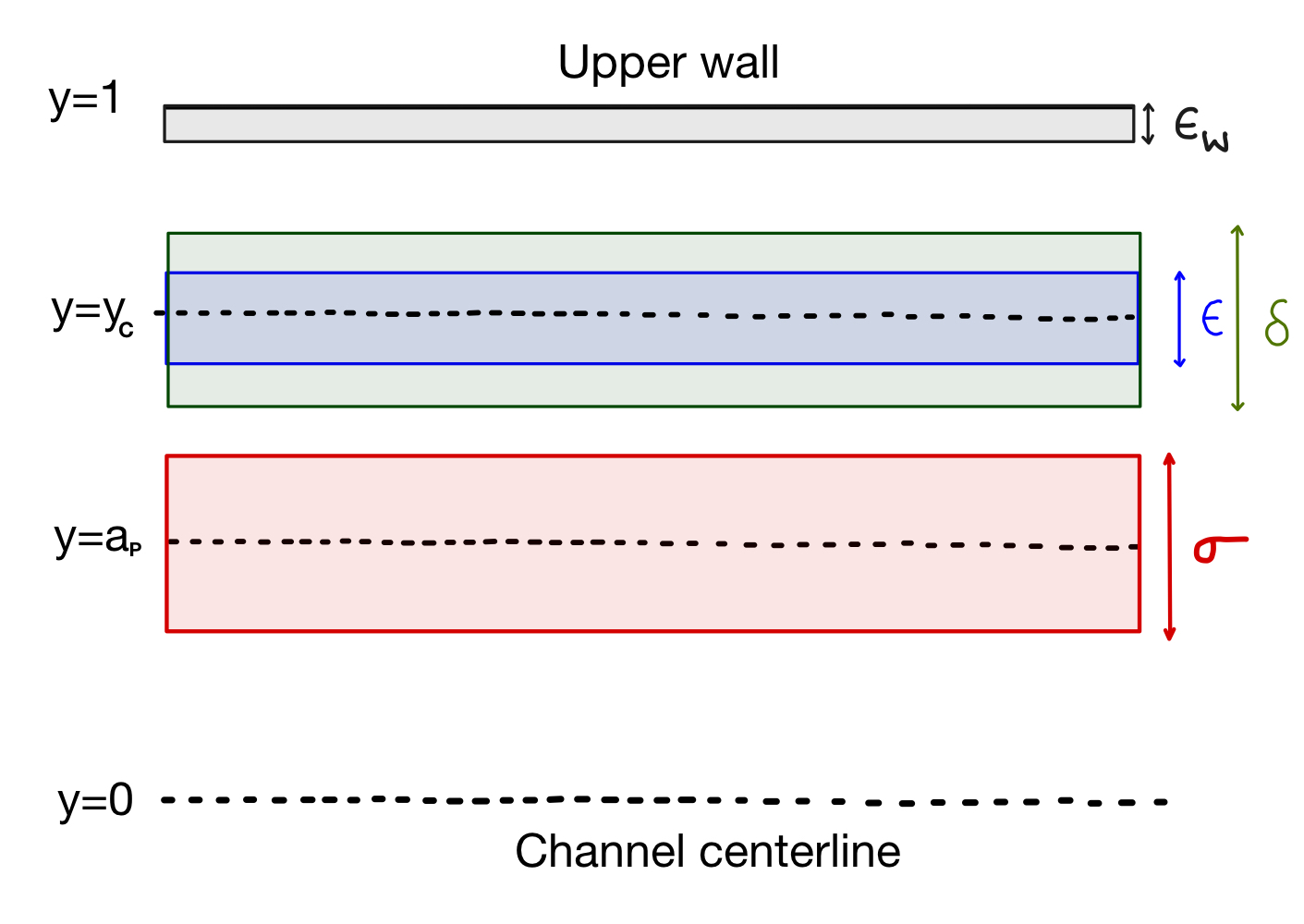}
         \caption{}
         \label{fig:crit_non_overlap}
         \end{subfigure}
         \begin{subfigure}[b]{0.48\textwidth}
         \centering
         \includegraphics[width=\textwidth]{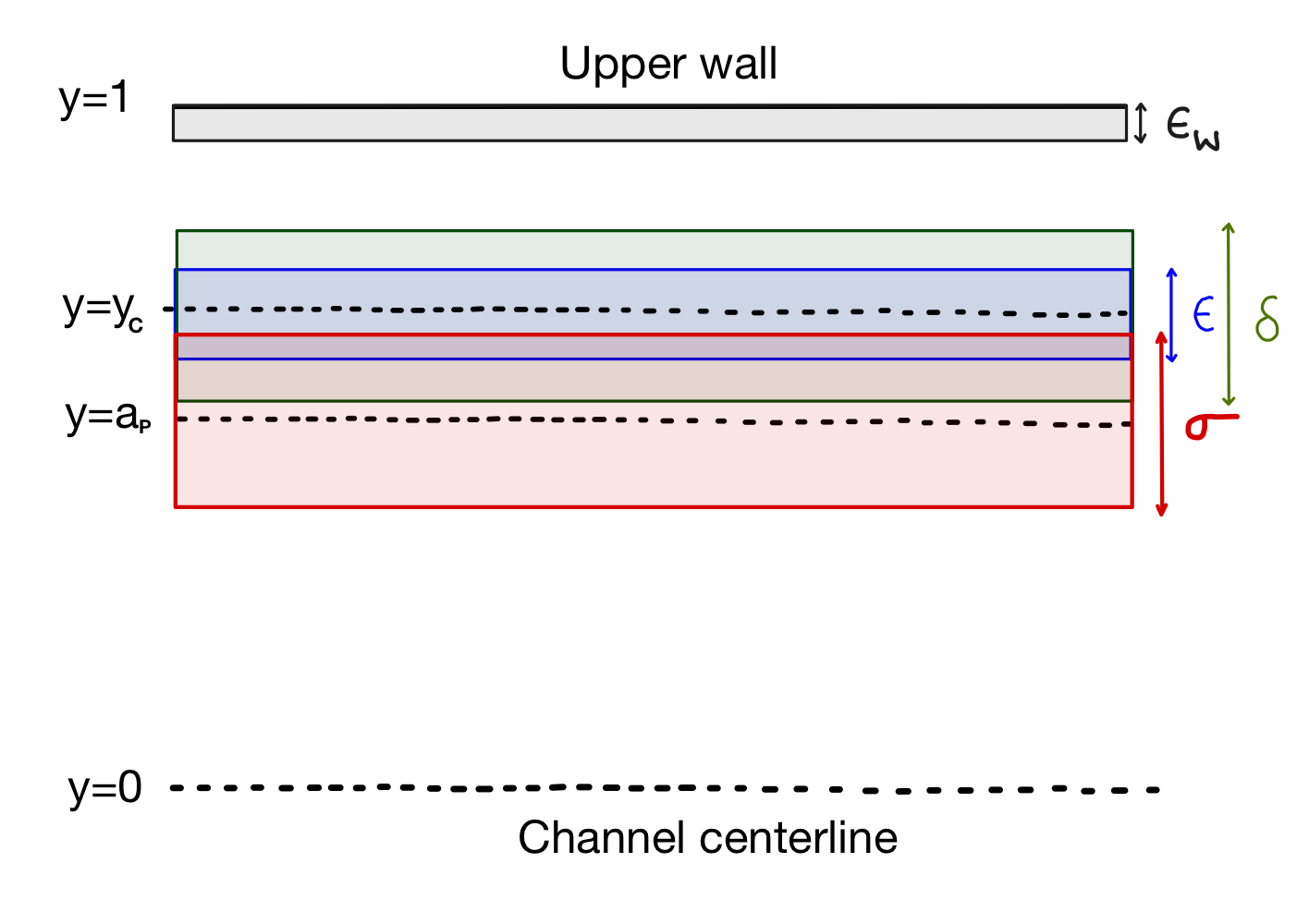}
         \caption{}
         \label{fig:crit_overlap}
         \end{subfigure}         
  \caption{Schematic of layers within which there are rapid variations in one or more physical quantities. The perturbation stream function $u_y$ and the perturbation suspension velocities $v_x$ and $v_y$ display critical layers of thickness $\epsilon$ and $\delta$ respectively around $y=y_c$. Additionally, the swift transition in the suspension mass fraction profile occurring at $y = a_p$ within a small region characterized by size $\sigma$ is seen. This depiction shows only the top half of the channel; the other half being symmetric. (a) condition where the layers are distinct, (b) overlap condition.}
\label{fig:crit}
\end{figure}

It is useful to begin this section by defining the critical layer, since the physics therein dominate this discourse. It is a relatively thin layer centered around the critical point $y_c$ in the channel, where the mean-flow velocity is the same as the phase speed of the dominant normal-mode perturbation, i.e., $U(y_c)=c$ \citep{lin1945stabilitya,lin1945stabilityb,lin1946stability_part3}. The particle layer on the other hand, as seen from equation (\ref{feqn}), is centered around $y=a_p$. If $y_c$ and $a_p$ are in close proximity, such that both layers overlap, we term it as the `overlap' condition, and when these layers are distinct and well-separated, we term it a `non-overlap' condition. The channel comprises the critical layer, the wall layer, the particle laden layer and the inviscid outer layer; and different physics can appear in each. The first three are shown schematically in figure \ref{fig:crit}, under overlap and non-overlap conditions. The need for asymptotic analyses in the different layers is motivated below.

In the dilute particle limit, whether or not the particles are far denser than the fluid, it can be worked out that the viscosity variation terms will not enter the dominant balance, so we may work with the single equation \ref{generalised_ev_prob}.

\subsection{Motivation}

\begin{figure}
         \centering
         \includegraphics[width=\textwidth]{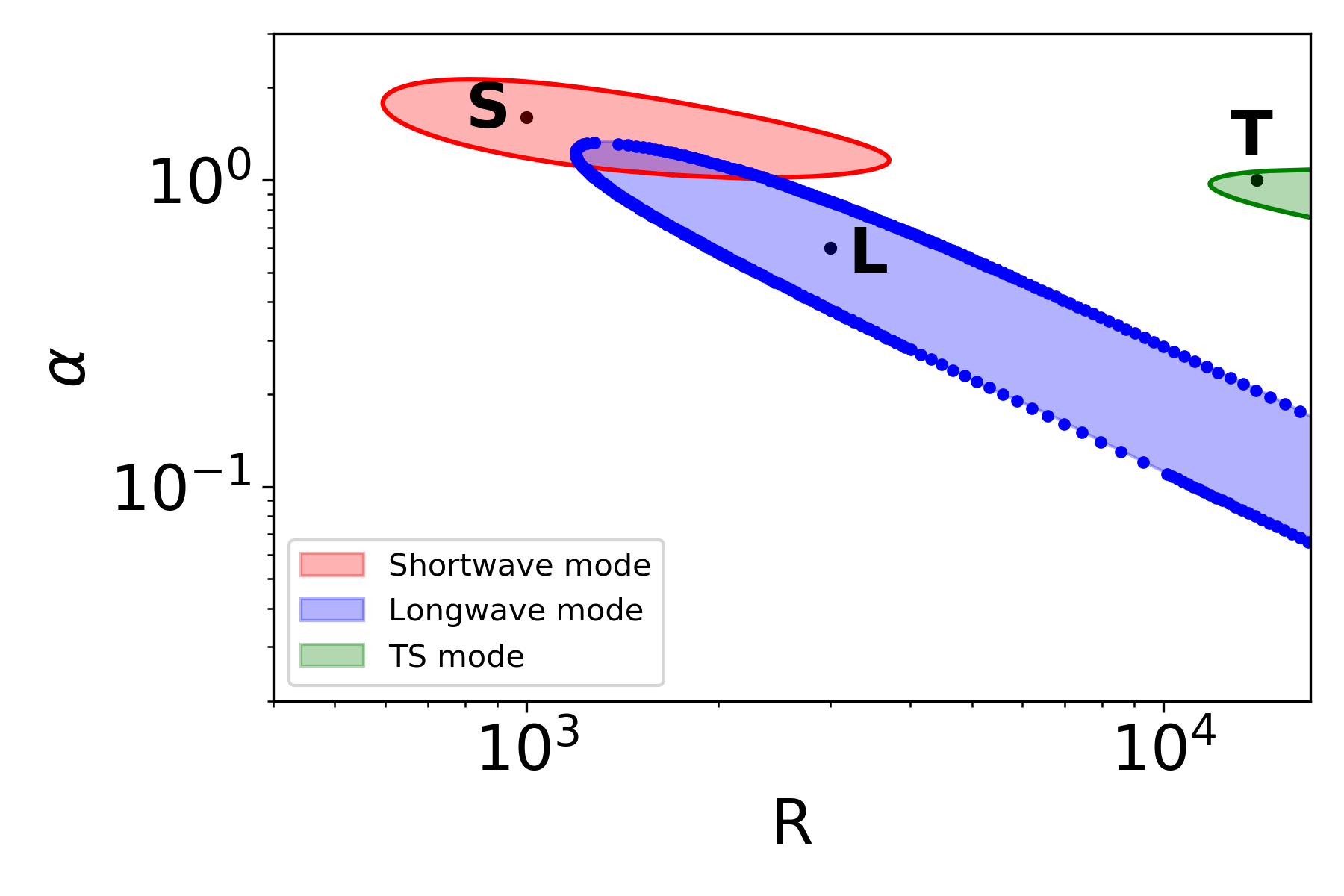}
     \caption{The three distinct modes of instability, shown by the shaded regions. A specific choice of parameters is made here, where overlap conditions prevail: the peak of the mean particle concentration profile has an amplitude $f_{max}=0.70$, and is positioned at $a_p=0.75$. The thickness of the particle-layer is $\sigma=0.1$ and the Stokes number is $S=8\times10^{-4}$. This figure is representative of a wide range of parameters under overlap conditions. The points marked S, L and T are representative of shortwave, longwave and Tollmien-Schlichting modes respectively, and will be elaborated on.}
     \label{example_three_modes}
     \end{figure}
     We saw in figure \ref{fig:validation_neutral_curves} that in the case of constant particle loading, the TS mode of instability is modified by particles. Even with non-uniform particle loading, under non-overlap conditions, the same is observed. On the other hand, under overlap conditions, the picture is very different, and an example is shown in figure \ref{example_three_modes}. Here the TS mode is seen as a minor blip on the right of the figure. Two other modes of instability are now seen, which occur at much lower Reynolds number. The fact that these modes are distinct from the TS mode is evident from the separate regions in $\alpha-R$ space they occupy. To distinguish between them, the two lower Reynolds number modes of instability will be termed shortwave and longwave respectively, while remembering that the so-called shortwave mode actually has perturbation wavelengths of $O(1)$, i.e.,  comparable to the channel width (a wavelength of $\alpha=1$ is $2\pi$ times the half-width). The longwave modes extend from $O(1)$ 
 to far lower wavenumbers. The shortwave mode of overlap instability occurs over the smallest Reynolds numbers, ranging from a few hundreds to a few thousands, while the longwave mode spans decades in the Reynolds number, with the instability Reynolds number and the typical wavelength increasing together. In the following section we show that both of these are overlap modes of instability, caused by the variation of particle concentration within the critical layer. In Appendix B we perform a similar analysis for the wall layer.
 
In explaining the mechanism for the low Reynolds number instabilities, we may pursue one of two approaches. For both of them, we must begin by deriving the dominant balance in the critical layer. Once we have the lowest order equations in the critical layer, we could solve the equations in the inner (critical) layer, and perform a matching with the outer layer (inviscid) solutions to obtain the full solutions. But this would yield no extra information, since we can already solve the full solutions. We therefore follow a second approach: of writing down a minimal composite theory for particulate shear flow. This theory \citep{narasimha2000minimal,govindarajan2001estimating,bhattacharya2006critical} will obtain a reduced set of equations describing the stability problem. The reduced equations will contain all terms in the complete stability equations (\ref{minimal_equation}) which participate in the dominant balance somewhere in the flow, and none of the terms which do not participate in this anywhere.

\subsection{Dominant balances in the critical layer}

We first summarise existing knowledge in the context of a clean fluid, and then derive dominant balances within the critical layer in particulate shear flow. 

In the Orr-Sommerfeld equation (\ref{os}) for a clean fluid, it is seen that the highest, i.e., fourth-order, derivative term in $y$ is scaled by the inverse of the Reynolds number. Now even if the Reynolds number approaches infinity, this term may not be dropped, because if it is, we will not be able to satisfy all four boundary conditions associated with equation (\ref{os}). This is thus a classical singular perturbation problem \citep{vanDyke1964}, where the highest derivative term becomes as big as the terms on the left-hand side in some portions of the flow.  There are two layers \citep{lin1945stabilitya, lin1945stabilityb, lin1946stability_part3} where viscous effects are important and gradients are large: the wall layer, of thickness $\epsilon_w \sim R^{-1/2}$, and the critical layer, of thickness $\epsilon \sim R^{-1/3}$ where, as defined above, $U \sim c$. It is the latter which is of primary interest to us to explain the mechanism of the overlap instabilities.
To perform a similar analysis for particulate shear flow, we limit ourselves here to the regime where $RS \sim O(1)$, which is reasonable for dilute particle suspensions at high Reynolds number. A similar analysis may be carried out for any order of magnitude of this quantity. There are three layers we pay attention to on each side of the centreline, and these are shown in Fig. (\ref{fig:crit}) for one half of the channel. There is also a wall layer shown, which will be discussed separately in Appendix B. There are now two critical layers: for the fluid and for the particle flow, of thickness $\epsilon$ and $\delta$ respectively, and the layer where the particles are concentrated, of thickness $\sigma$, which is pre-specified. The scales $\epsilon \ll 1$ and $\delta \ll 1$ are as yet unknown, and will be determined below. Fig. (\ref{fig:crit_non_overlap}) is a schematic for conditions where the particle layer and the critical layer are distinct, which we shall refer to as the non-overlap condition, and Fig. (\ref{fig:crit_overlap}) depicts the overlap condition.  
 
We derive equations within the critical (inner) layer in the inner variables $\xi$ and $\lambda$, defined as
\begin{equation}
    \xi=\frac{y-y_c}{\epsilon},\quad {\rm and} \quad \lambda=\frac{y-y_c}{\delta}
    \label{expan1}
    \end{equation}
and will select $\epsilon$ and $\delta$ to ensure that the derivatives of the fluid velocity components in $\xi$ and the particle velocity components in $\lambda$ are $O(1)$. In addition, it is useful to define
\begin{equation}
   \chi=\frac{y-a_p}{\sigma}.
    \end{equation}
    
To derive the dominant balances we write the relevant variables in the form of series expansions within the critical layer as 
\begin{equation} 
 u_y=\sum_{n=0}^{\infty} \epsilon^n  u_{y, n}(\xi), \quad v_y  =\sum_{n=0}^{\infty} \delta^n v_{y,n}(\lambda) \quad {\rm and} \ \  v_x  =\sum_{n=0}^{\infty} \delta^n v_{x, n}(\lambda).
\label{expansions}
\end{equation}
In this layer the mean flow may be written in the following expansion:
    \begin{equation} 
 U(y)-c=(y-y_c)U_{c}^{'}+\frac{(y-y_c)^2}{2\!} U_{c}^{''}+\dotsb.
\label{Uexpansion}
\end{equation}
The relative magnitudes within the critical layer of the two components of flow can be established from the continuity equation. We have
\begin{equation}
\sum_{n=0}^{\infty}\epsilon^n \left[i \alpha u_{x,n} + \frac{1}{\epsilon} D_\chi u_{y,n}\right] = 0.
\end{equation}
Constructing hierarchies of equations of different powers of $\epsilon$ yields 
\begin{equation}
    u_{y,0}=0
    \label{uy_0}
\end{equation} 
and shows that the coefficient of a particular power of $\epsilon$ in $u_x$ is related to that which is one order higher in $u_y$. This is in fact a natural consequence of incompressibility. 
For the particle field, from the equation (\ref{visc_para_definitions}), and using equations (\ref{expansions}) and (\ref{Uexpansion}), we get
\begin{equation}
    \left[ U-c -\frac{i}{\alpha SR}\right]v_x=i\frac{U^{\prime}}{\alpha}v_y+\frac{1}{\alpha^2  SR}Du_{y}
    \label{vx_AX_Bx}.
\end{equation}
At the next two orders, using equation (\ref{vx_AX_Bx}) as well as the incompressibility condition $D_\xi  u_{y, 1} =- i \alpha u_{x,0}$ in the critical layer, we get
    \begin{equation}
v_{x, 0} = u_{x, 0}, \quad
    v_{x,1}=\frac{\epsilon}{\delta} \left\{\frac{i}{\alpha}D_{\xi}u_{y,2} -i\alpha SRU_c^{'}\xi v_{x,0} \right\}-SRU_{c}^{'}v_{y,1}.
    \label{vx_1}
\end{equation}
This yields $\delta \sim \epsilon$, and without loss of generality, we choose $\delta = \epsilon$. The critical layer thickness as perceived by the fluid and the particles is thus identical. 

Using the third row of the matrix equation (\ref{generalised_ev_prob}) along with equations (\ref{expansions}) and (\ref{Uexpansion}), and collecting terms at the lowest order in the expansion, we obtain
\begin{equation}
v_{y, 0} = u_{y, 0},
 \label{vy_0}
\end{equation}
which we know to be $0$ from equation (\ref{uy_0}). In other words, the expansions for the normal velocity components for the particles too begin one order higher than the streamwise component. At the next two orders, from the equation (\ref{visc_para_definitions}) 
\begin{equation}
    \left[ U-c -\frac{i}{\alpha SR}\right]v_y=-\frac{i}{\alpha SR}u_{y}
    \label{vy_AX_Bx}
\end{equation}
we get 
\begin{equation}
    v_{y,1}=\ u_{y,1}, \quad {\rm and} \quad v_{y,2}=u_{y,2} -i\alpha SR U_c^{'}\xi v_{y,1}.
    \label{vy_12}
\end{equation}
From equations (\ref{expansions}), (\ref{vx_1}), and (\ref{vy_12}), we can see that the components of $\mathbf{v}$ and $\mathbf{u}$ differ from each other only at order $\epsilon$ relative to their largest value in the critical layer. This is consistent with the expectation that the particle velocity field must closely follow the fluid velocity field for low Stokes numbers. This analysis yields a measure of the difference between the two. 

Finally, we may derive the dominant balance for fluid velocity in the critical layer from the equation (\ref{visc_para_definitions}), along with (\ref{expansions}) and (\ref{Uexpansion}), and the 
Taylor expansion 
\begin{equation}
    \bar{f}= 
    \begin{cases}
   \bar{f}_{c}+(y-y_{c})\bar{f}_{c}^{\prime}+\frac{(y-y_c)^2}{2}\bar{f}_{c}^{\prime\prime}+\dotsb, & \text { overlap case } \\ 
    \bar{f}_c, & \text { non-overlap case }
    \end{cases}
\end{equation}
and under overlap conditions we may rewrite this in the relevant variable $\chi$.  
Now there are different choices possible for the small parameters. We therefore apriori retain all terms which may participate in the dominant balance, and after some algebra obtain the following composite lowest-order equation
\begin{equation}
   \left[(1+\bar{f}) \xi U_{c}^{\prime}D_{\xi}^{2}+i\frac{1}{\alpha R \epsilon^3}D_{\xi}^{4}-\frac{\epsilon}{\sigma}U_{c}^{\prime}(D_{\chi}\bar{f})(I-\xi D_{\xi})\right]u_{y,1}=0,
    \label{composite}
\end{equation}
where $I$ is the identity operator. We will have one of the following four distinct cases arising. Case 1 is the non-overlap case, while the others are for overlap conditions.\\
\textbf{Case-1:} The particle layer and critical layer are well separated, or $1\sim\frac{1}{\alpha R\epsilon^3}\gg \frac{\epsilon}{\sigma}$, i.e., the size of the particle layer significantly exceeds that of the critical layer. The third term in equation (\ref{composite}) now becomes negligible and in both cases this equation simplifies to:
\begin{equation}
   \left[-iU_{c}^{\prime} \xi D_{\xi}^{2}+D_{\xi}^{4}\right]u_{y,1}=0.
    \label{composite_case1}
\end{equation}
The balance, with $\epsilon={\alpha R}^{-1/3}$, is identical to the case with no particles, for the following reasons. When the particle layer and critical layer are well separated, the mass fraction $\bar {f}$ and its derivative ($D_\chi \bar {f}$) / $\sigma$ practically negligible or much less than $O(1)$ in the critical layer, regardless of how small the particle layer $\sigma$ is. In the other scenario, of a thick particle layer, we have or $1\sim\frac{1}{\alpha R\epsilon^3}\gg \frac{\epsilon}{\sigma}$. 

$\textbf{Case-2:} \quad 1\ll\frac{1}{\alpha R\epsilon^3}\sim \frac{\epsilon}{\sigma}$, i.e., the particle-laden layer is markedly smaller than the critical layer. The scaling that emerges is $\epsilon \sim (\frac{\sigma}{\alpha R})^{1/4}$, and upon replacing $\sim$ by equality, equation (\ref{composite}) simplifies to:
\begin{equation}
  \left[ D_{\xi}^{4}+i\left\{(D_{\chi}\bar{f})_{c} U_{c}^{\prime}\right\}(I-\xi D_{\xi})\right]u_{y,1}=0.
    \label{composite_case2}
\end{equation}
The variation in particle concentration is as important as the largest viscous effects in the critical layer.
\\
$\textbf{Case-3:} \quad 1\sim \frac{\epsilon}{\sigma} \gg\frac{1}{\alpha R\epsilon^3}$. The size of the particle layer is comparable to that of the critical layer, and the critical layer is much wider than that dictated by the scaling on the inverse Reynolds number. This case is a mathematical possibility, but is unlikely to occur physically, since the critical layer at large Reynolds will normally be influenced by the Reynolds number, and become thinner as Reynolds number increases. Under these hypothetical conditions, viscous effects appear only at higher order, since equation (\ref{composite}) simplifies to:
\begin{equation}
  \left[(1+\bar{f}) \xi U_{c}^{\prime}D_{\xi}^{2}-(D_{\chi}\bar{f}) U_{c}^{\prime}(I-\xi D_{\xi})\right]u_{y,1}=0,
    \label{composite_case3}
\end{equation}
where we have set $\epsilon=\sigma$. 
\\
$\textbf{Case-4:} \quad 1\sim \frac{\epsilon}{\sigma} \sim\frac{1}{\alpha R\epsilon^3}$. The size of the particle layer is comparable to that of the critical layer, and viscosity plays a significant role as well. Equation (\ref{composite}) now becomes
\begin{equation}
   \left[U_{c}^{\prime}(1+\bar{f}) \xi D_{\xi}^{2}+i D_{\xi}^{4}-\kappa(D_{\chi}\bar{f}) U_{c}^{\prime}(I-\xi D_{\xi})\right]u_{y,1}=0,
    \label{composite_case4}
\end{equation}
Here, we define $\epsilon=(\alpha R)^{-1/3}$ and $\frac{\epsilon}{\sigma}\equiv \kappa$. 

Equation (\ref{composite_case4}) completely describes the critical layer for the thickness $\sigma$ we consider. 
Cases 2 to 4 correspond to overlap conditions, and the variation of the particle concentration within the critical layer, estimated by $D_{\chi}\bar{f}$ is an important player in the critical-layer balance, altering it fundamentally. We have thus established that the concentration profile will alter the fundamental nature of shear flow instability, only under overlap conditions. This effect would be absent with uniform particle loading, where only under case 4, we will have a factor $(1+\bar{f})$ which merely rescales the Reynolds number. 

\subsection{Construction of the minimal composite theory}
The lowest order equation in the critical layer is given by equation (\ref{composite_case4}) which includes all the effects seen in (\ref{composite_case2}) as well. The lowest order equation in the wall layer is derived in Appendix B, and is is given by 
\begin{equation}
    \left[-c\left\{1+ \frac{\bar{f}}{1-i\alpha cSR}\right\} D_{\xi}^{2} + i D_{\xi}^4 - \kappa_w\left\{\frac{cD_{\chi}\bar f}{1-i\alpha cSR}\right\}D_{\xi}\right]u_{y,1} = 0,
    \label{wall_layer_lowest}
\end{equation}
where $\kappa_{w}=\epsilon_{w}/\sigma$. In the rest of the channel, a particle-laden counterpart of the Rayleigh equation, where all viscous effects are neglected in the stability operator, is valid. We now construct a reduced equation which includes every term in the complete equations (\ref{generalised_ev_prob}) from which any of the dominant terms in the three layers originates, and neglects all other terms.

The final minimal composite equation is 
\begin{equation} \left[(U_{*}-c)(D^2-\alpha^2)-U''\right]\psi-(J'f')\psi+(Jf')D\psi=\frac{1}{i\alpha R}D^4 \psi, 
\label{minimal_equation}
\end{equation} 
with, as before, 
\begin{equation} 
U_{*} \equiv U+J\bar{f} \quad \quad J\equiv \frac{U-c}{1+i\alpha (U-c)SR}. 
\end{equation}
The terms that are not included in the above equation compared to the full equation (\ref{generalised_ev_prob}) are: $-J^{''}\bar{f}\psi -(1/i\alpha R)(-2\alpha^2D^2+\alpha^4)\psi$, apart from all the viscosity stratification effects which vanish from the minimal physics in a dilute suspension. This is because the derivative of the mean viscosity is $O(f_{max}/[\gamma\sigma])$, which is small for a dilute suspension. Moreover, since $J$ is a function of $y$, we see that (\ref{minimal_equation}) represents a significant reduction of the complete stability operator. It comprises the inviscid stability operator of Rayleigh and the highest order derivative in the viscous operator. For a constant particle loading, the only effect due to particles would come from the modified effective mean flow profile $U_*$. Besides these, terms appear are due to variations in the particle concentration profile which are critical. Note importantly that the minimal equation does not reduce to the Orr-Sommerfeld equation in any limit. 
     
It remains to be seen whether the essential physics is contained in the minimal composite equation. 
\begin{figure}
\includegraphics[width=1.0\textwidth]{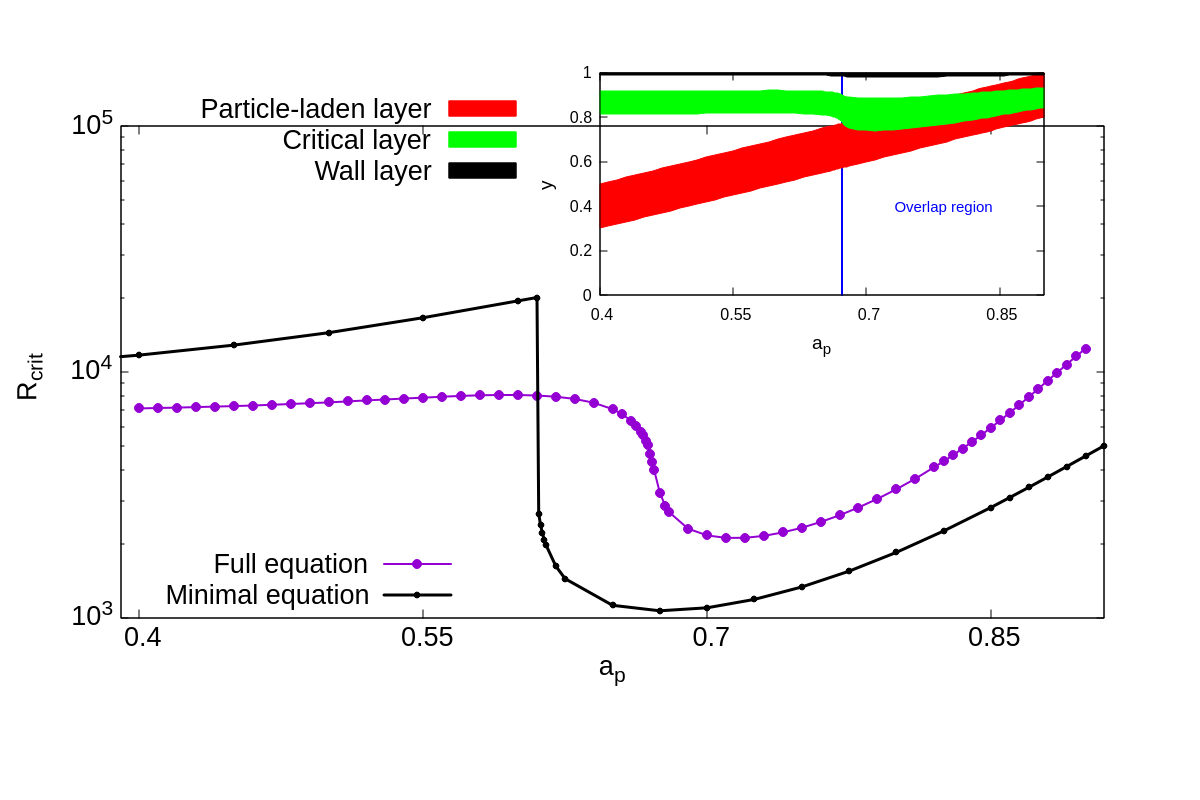}
  \caption{With a specified amplitude of $f_{max}=0.1$, a peak width of $\sigma=0.1$ for the mass fraction profile, and a Stokes number of  $S=2.5\times 10^{-4}$,
The purple curve and the black curve illustrate the critical Reynolds number as a function of the position of the peak, $a_p$ of the mass fraction profile for the full equation (\ref{generalised_ev_prob}) and the minimal equation (\ref{minimal_equation}) respectively. The green, black and red bands in the inset illustrate the notional critical layer, the wall layer and the particle-laden layer, respectively, as they vary with $a_p$. }
\label{fig:a0_plot}
\end{figure}
The best parameter to make this explicit is the location, $a_p$, of the maximum in particle concentration. In figure \ref{fig:a0_plot} we show how the critical Reynolds number (the lowest Reynolds number for instability), $R_{crit}$, changes with $a_p$. The purple line with particles represents the full solution to equation (\ref{generalised_ev_prob}), whereas the black line is the solution of the minimal composite equation (\ref{minimal_equation}). As $a_p$ increases, the particle-laden layer shifts towards the wall. We see that below $a_p \sim 0.6$, there is a continuous increase in the critical Reynolds number. But beyond this, we see a sudden and large drop in $R_{crit}$, going down to values less than half of that a clean channel ($R_{crit}=5772.2$). At large $a_p$, however, i.e., when the particle-laden layer is very close to the wall, the trend is reversed again and a large stabilization is seen. The complete trend is captured by the minimal composite equation, although, as is to be expected, the agreement with the full solution is only qualitative. 

The sensitivity of the critical Reynolds number with the location of the particle-laden layer is now seen to have its root in the dynamics within the critical layer of the dominant disturbance.   
The critical layer is shown in the inset of figure \ref{fig:a0_plot}, as a function of $a_0$. A notional thickness of  $R^{-1/3}$ is shown in this sketch. Shown in the same figure is the linear movement of the particle-laden layer with $a_p$. The large changes in stability occur in the regime of $a_p$ when the two layers overlap. A major portion of the disturbance kinetic energy is known to be produced within the critical layer in clean channel flow. We shall see that this is true of particulate flow too. Note that the wall layer (also shown in the figure) is unimportant in the dominant balance.

\subsection{Summary of instability features in the overlap and non-overlap contitions}
     
\begin{figure}
     \centering
     \begin{subfigure}[b]{0.7\textwidth}
         \centering
         \includegraphics[width=\textwidth]{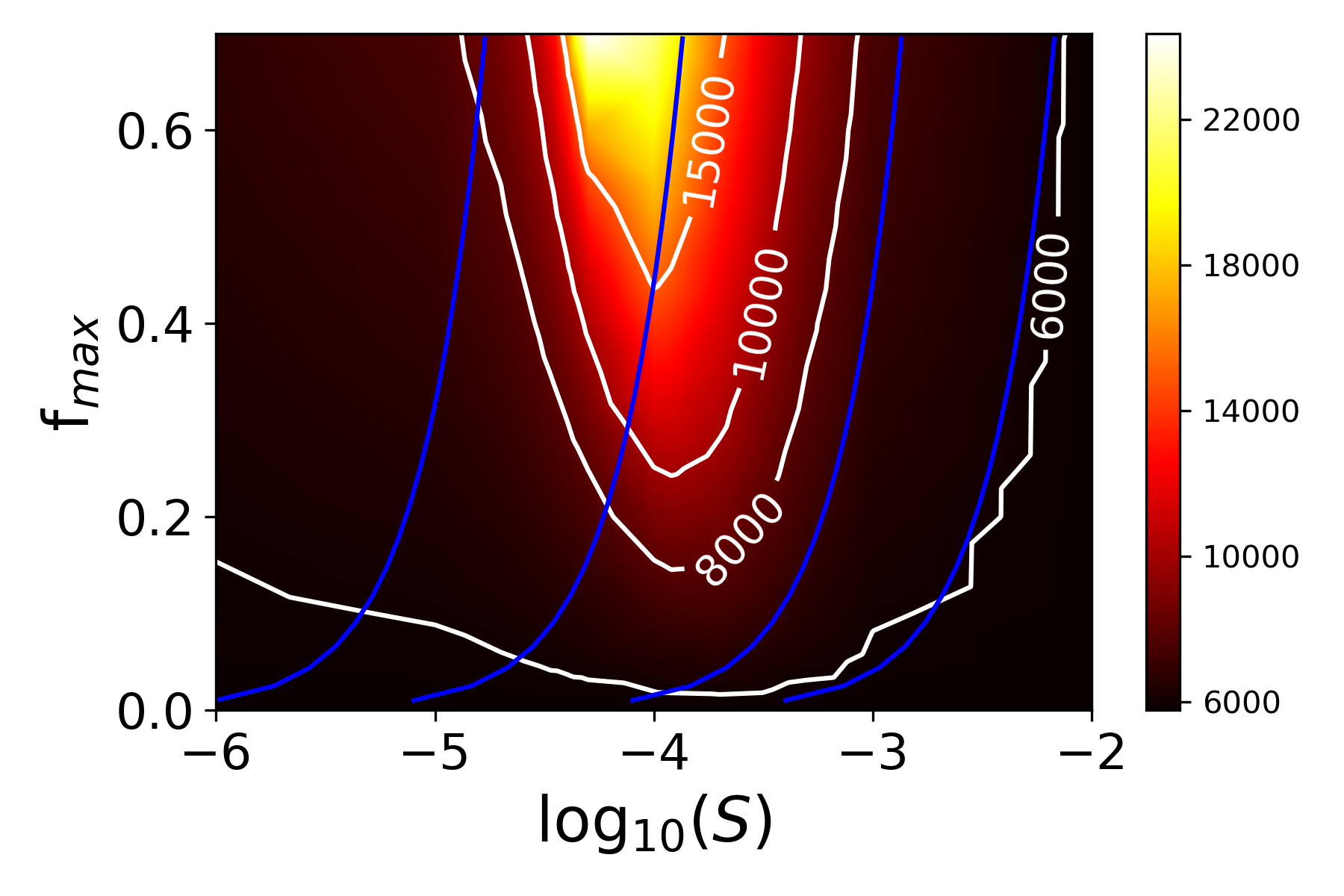}
         \caption{}
         \label{fig:non-overlap}
     \end{subfigure}
         \begin{subfigure}[b]{0.7\textwidth}
         \centering
         \includegraphics[width=\textwidth]{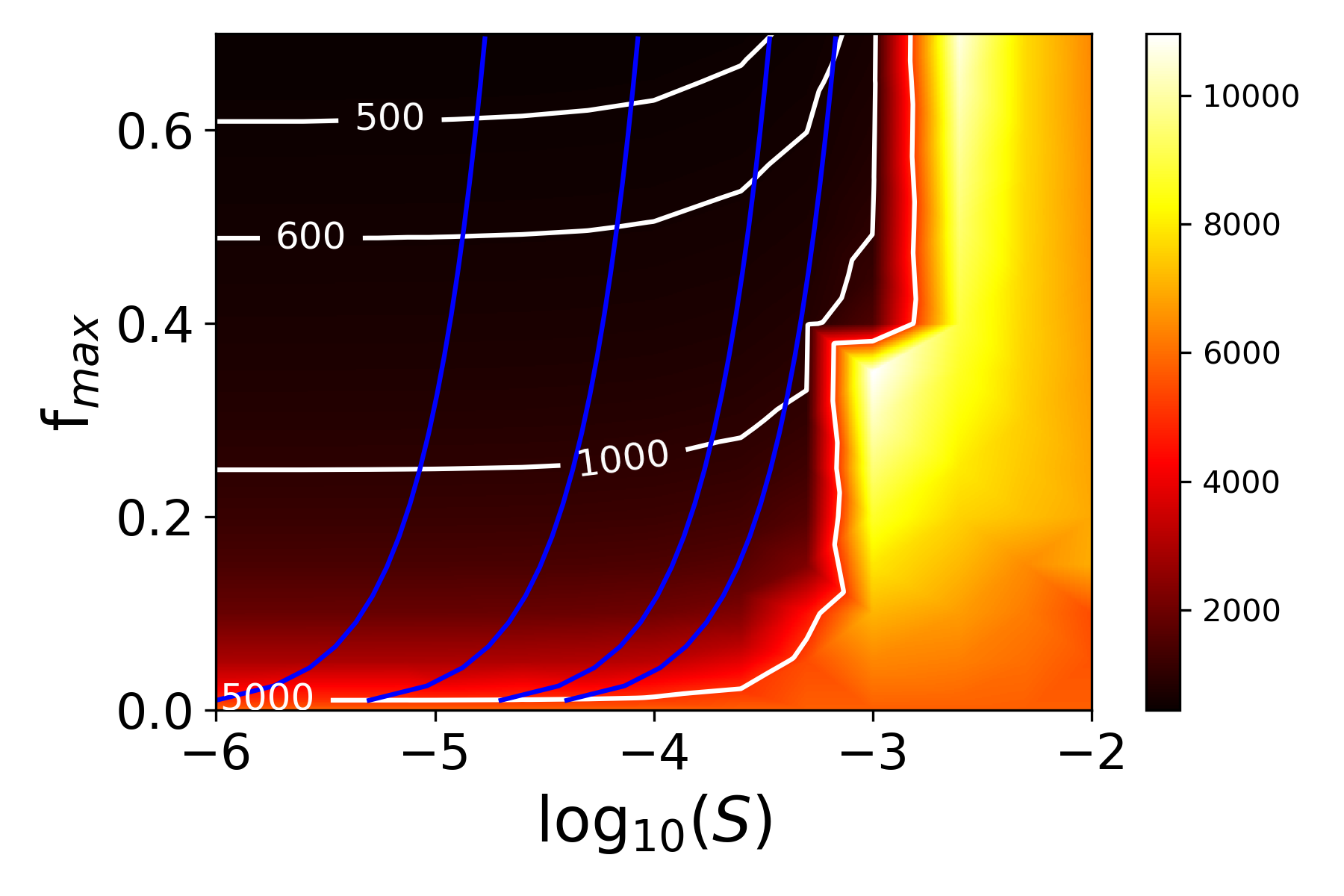}
         \caption{}
         \label{fig:overlap}
     \end{subfigure}
     \caption{Phase plot of the critical Reynolds number, shown in colour, as a function of the Stokes number $S$ and the particle loading strength $f_{max}$. (a) A case where there is no overlap mechanism in operation, with $a_p=0.40$, and (b) where it is in operation, with $a_p=0.75$. Note the difference in the colourbars in the two figures. In both figures $\sigma=0.1$, and the blue lines represent curves of constant particle number density $N$. The value of $N$ decreases from left to right, with the non-dimensional quantity $9\sqrt{2}H^3N\sqrt{\rho_{f}/\rho_p}$ being $[1.00\times10^7, 4.42\times10^5, 1.40\times10^4, \ {\rm and } \ 1.25\times10^3]$ for plot (a), and $[1.00\times10^7, 8.94\times10^5, 1.12\times10^5, \ {\rm and } \ 3.95\times10^4]$ for plot (b). }
     \label{general}
\end{figure}
Before we discuss the energy budget, we summarise some additional features of the modes of instability.
The maximum, $f_{max}$, in the particle concentration and the thickness, $\sigma$, of the particle-laden layer, have a quantitative, rather than qualitative, effect. We fix $\sigma$ at $0.1$. 

Figure \ref{general} summarises the variation of the critical Reynolds number (given in colour with contour lines) with the Stokes number and the particle loading. We examine two situations: at $a_p=0.4$, where the overlap mechanism is not in operation, and at $a_p=0.75$, where it is. It is evident that both in quality and quantity the two situations are very different. Under non-overlap conditions, i.e., where the particle-laden layer lies in a different part of the channel from the critical layer (figure \ref{general}(a)), we see stabilization as particle loading is increased. The stabilisation is enormous in some portions of the regime, with the critical Reynolds number being extremely sensitive to either the particle loading or the Stokes number or both. The effect is largest at moderate particle Stokes number and is non-monotonic in the Stokes number. We now turn to figure \ref{general}(b), where completely the opposite trend is seen in response to increase in loading. For small to moderate Stokes number, with increase in loading, the flow is highly destabilised, with a sharp drop in critical Reynolds number. At high Stokes number, we see a reduction in the effect of particle loading, and a stabilization this time. The reason for this non-monotonicity could be as follows. Holding fixed the mass fraction $f_{max}$ of particles, as we increase the Stokes number, we are increasing the size of individual particles (from equation \ref{nondim} we see that the particle radius $r$ scales as the square-root of the Stokes number) and therefore reducing the number of particles. Thus, beyond a certain Stokes number, the forcing of the fluid by the particles comes down, and so does the effect of particle loading. We can check the consistency of this argument by examining the effect of Stokes number and mass loading on stability while holding the number density $N$ in equation (\ref{mass_frac_ex}) constant. Representative lines of constant $N$ are shown in figure \ref{general}. Following these lines from low $S$ and $f_{max}$ upwards, we again see different trends under overlap and non-overlap conditions. When the critical and particle-laden layers are distinct (figure \ref{general}(a)), and the particle number density is low (the two blue lines to the right of the figure), under non-overlap conditions, the critical Reynolds number is practically insensitive to changes in Stokes number and loading, i.e., the lines of constant $N$ appear parallel to lines of constant $R_{crit}$. This indicates that at small particle number density the critical Reynolds number is practically a function of $N$. At higher number densities (the two blue lines to the left of the figure) we see a strong stabilising effect with increase in $S$ while $N$ is held constant. This is in sharp contrast to the stability response under overlap conditions, where we see that the effect of increasing Stokes number (and at the same time, mass loading) and constant $N$ is monotonically and strongly destabilising. 

\begin{figure}
        \begin{subfigure}{0.48\textwidth}
            \includegraphics[width=\textwidth]{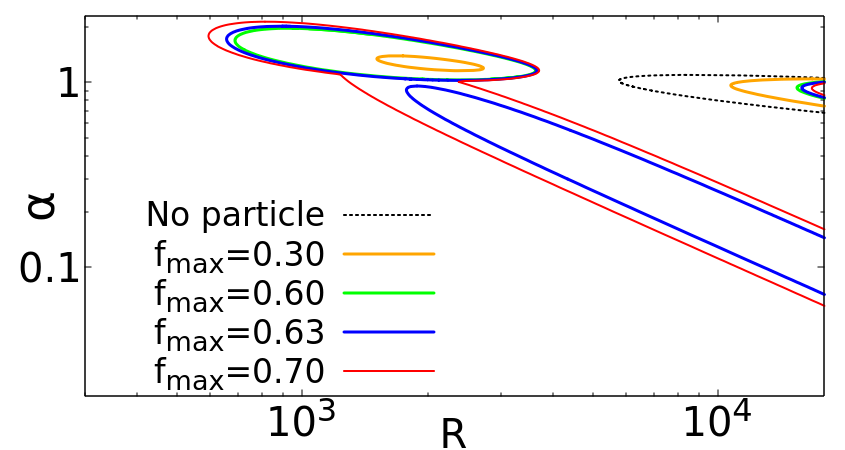}
        \caption{}
        \end{subfigure}
        \begin{subfigure}{0.48\textwidth}
         \includegraphics[width=\textwidth]{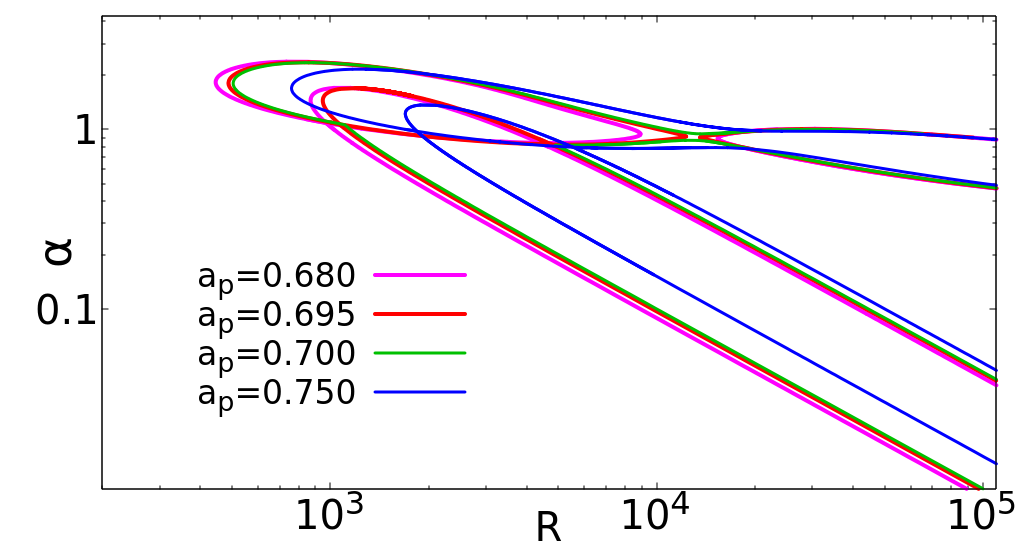}
         \caption{}
        \end{subfigure}
        
     \caption{(a) Stability boundaries for different amplitude of particle loading, with $a_p=0.75$ and $S= 8\times 10^{-4}$. 
     (b) Sensitive dependence of the stability boundaries on the location $a_p$ of the particle concentration peak. Here $f_{max}=0.4$ and $S=2.5\times 10^{-4}$. The shortwave overlap mode undergoes a merger with the TS mode just past $a_p=0.695$. Also there is significant intersection between the longwave and shortwave modes.}
     \label{three_mode_sensitivity}
\end{figure}
Figure \ref{three_mode_sensitivity} shows the dependence of the neutral boundaries on $f_{max}$ and $a_p$. A similar figure appears in \cite{rudyak1996stability} as well. The longwave mode is absent for smaller particle loading. The longwave mode is odd in $u_y$ whereas the shortwave and TS modes are even. The two even modes can go through merging bifurcations, as seen in figure \ref{three_mode_sensitivity}(b) at $a_p \sim 0.695$. After merger it is not easy to distinguish the boundary between the TS and the shortwave modes. The entire range of $a_p$ shown in this figure is small, underlining the sensitivity of the stability boundaries to this parameter. The even and odd overlap modes do not merge, but instead show a region of intersection, while each mode retains its character. They may always be distinguished by stipulating for the desired centerline conditions in the numerics. The longwave instability is particularly interesting because in channel flows, the most unstable perturbations are widely believed to be those which are even, with a maximum at the centreline, in the normal perturbation $u_y$. This assumption is so widespread that instability computations are often performed in the half-channel by imposing this symmetry at the centreline. Note that we use the terminology `odd' mode going by the normal perturbation $u_y$.

\begin{figure}
     \centering
          \begin{subfigure}[b]{0.48\textwidth}
         \centering
         \includegraphics[width=\textwidth]{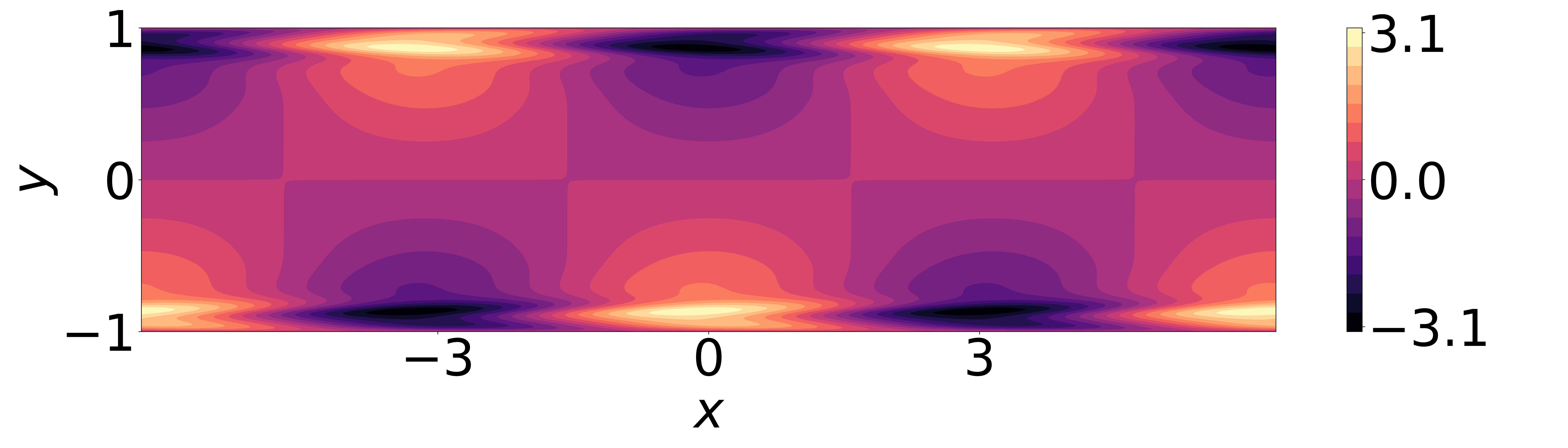}
         \caption{$\hat u_{x}(x,y)$}
         \label{fig:ux_f0d07_TS_mode}
     \end{subfigure}
     \begin{subfigure}[b]{0.48\textwidth}
         \centering
         \includegraphics[width=\textwidth]{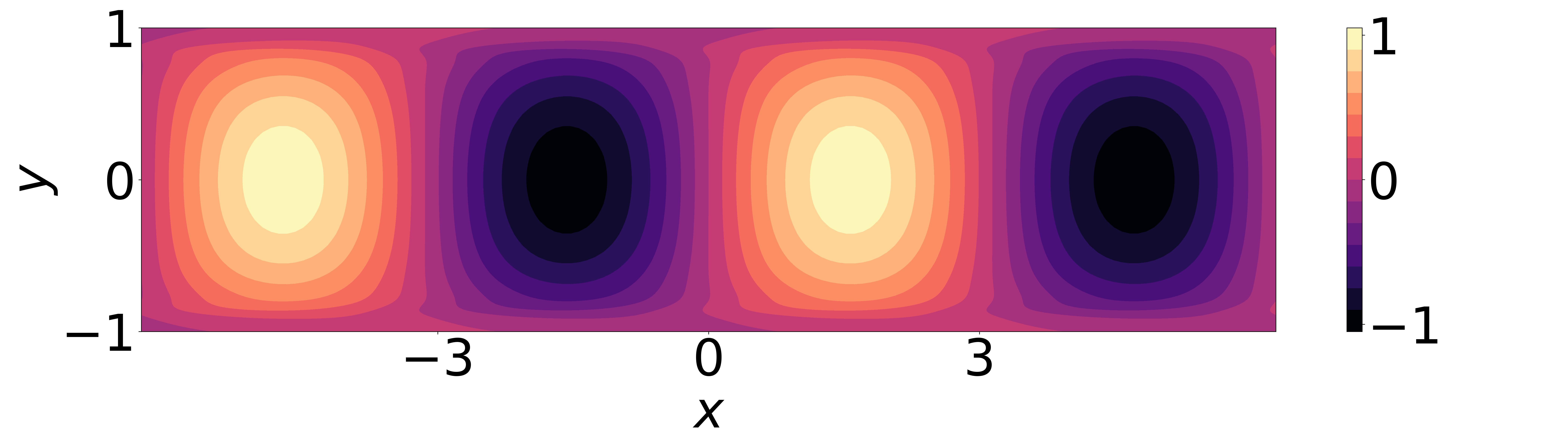}
         \caption{$\hat u_{y}(x,y)$}
         \label{fig:uy_f0d07_TS_mode}
     \end{subfigure}
    \hfill
        \begin{subfigure}[b]{0.48\textwidth}
         \centering
         \includegraphics[width=\textwidth]{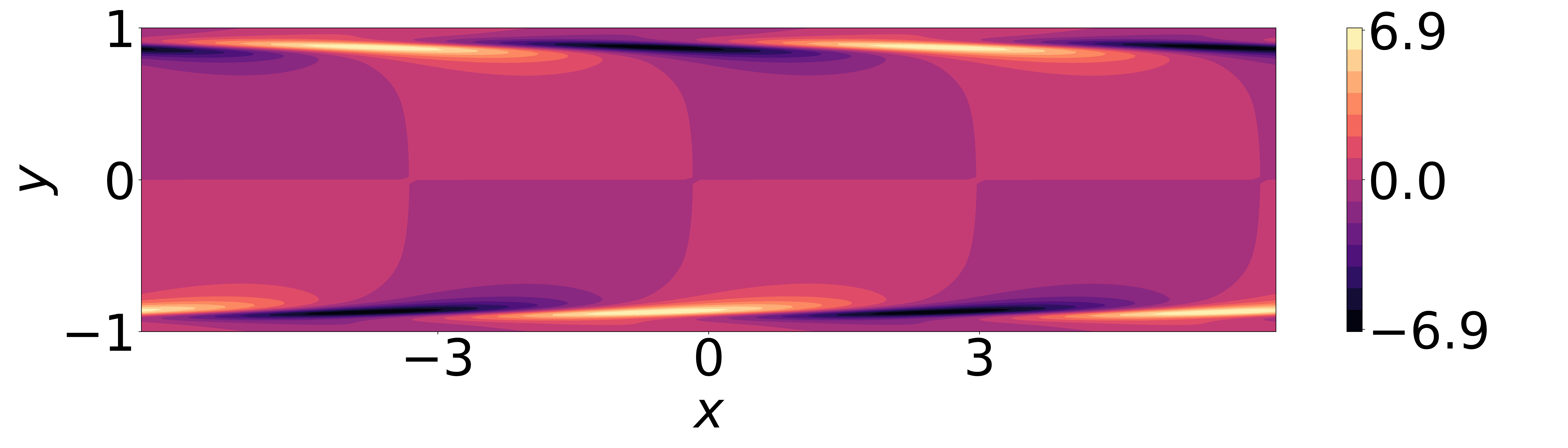}
         \caption{$\hat v_{x}(x,y)$}
         \label{fig:vx_f0d07_TS_mode}
    \end{subfigure}
    \begin{subfigure}[b]{0.48\textwidth}
         \centering
         \includegraphics[width=\textwidth]{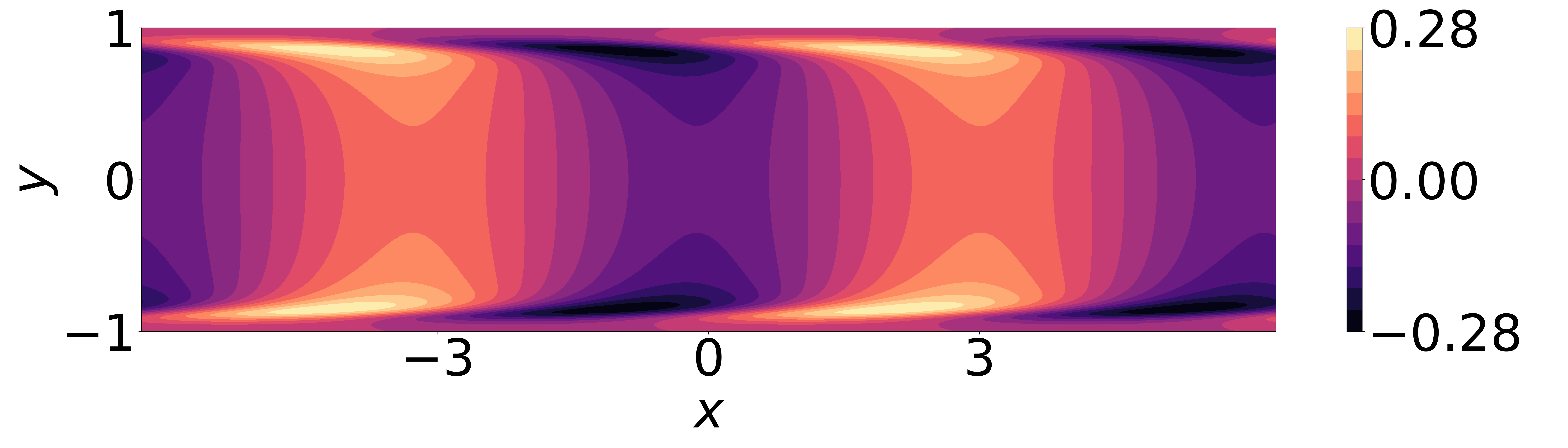}
         \caption{$\hat v_{y} (x,y)$}
         \label{fig:vy_f0d07_TS_mode}
    \end{subfigure}
     \caption{ Typical eigenfunctions of the TS mode in the $x-y$ plane. This mode depicts point `T' in figure \ref{example_three_modes} at $R=14000$, $\alpha=1.0$. A streamwise extent of two wavelengths is shown here and for all following eigenfunctions.}  
 \label{ei1}
     \centering
          \begin{subfigure}[b]{0.48\textwidth}
         \centering
         \includegraphics[width=\textwidth]{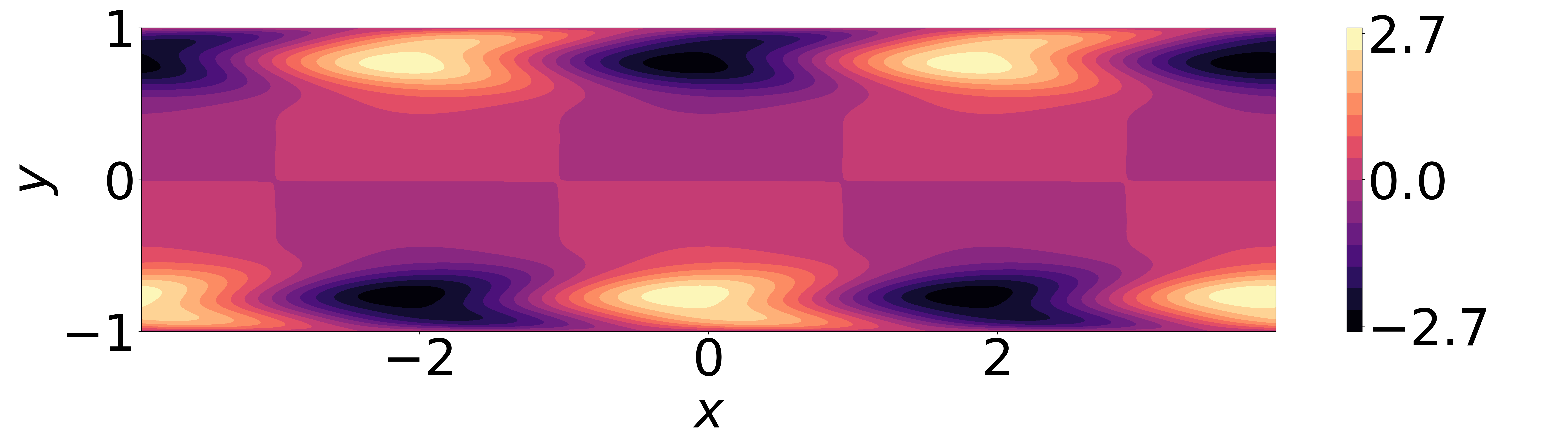}
         \caption{$\hat u_{x} (x,y)$}
         \label{fig:ux_f0d07_even_mode}
     \end{subfigure}
     \begin{subfigure}[b]{0.48\textwidth}
         \centering
         \includegraphics[width=\textwidth]{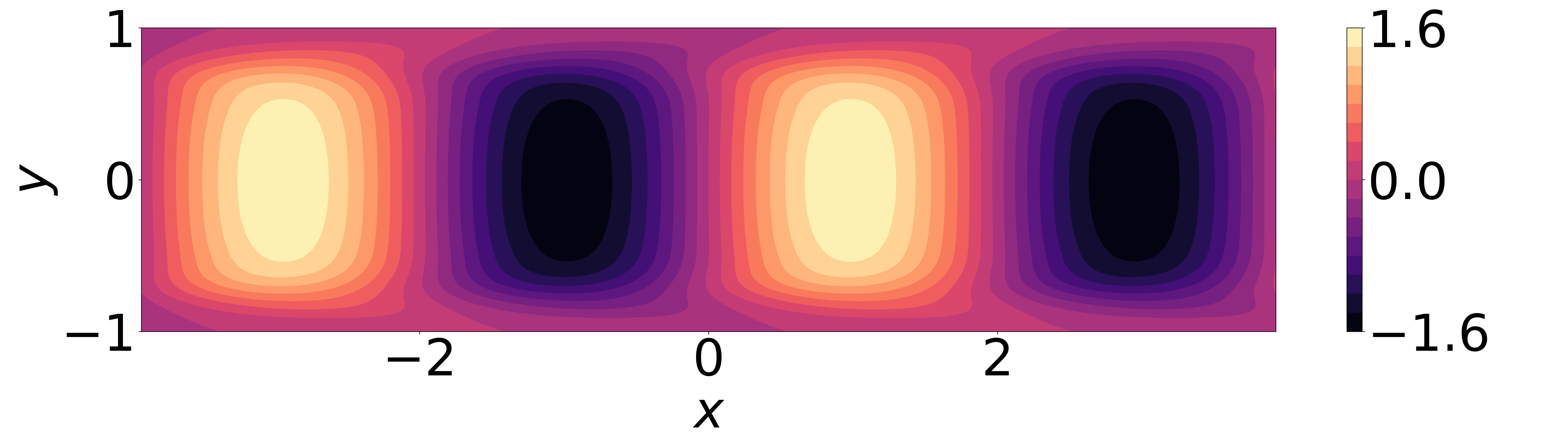}
         \caption{$\hat u_{y} (x,y)$}
         \label{fig:uy_f0d07_even_mode}
     \end{subfigure}
    \hfill
       \begin{subfigure}[b]{0.48\textwidth}
         \centering
         \includegraphics[width=\textwidth]{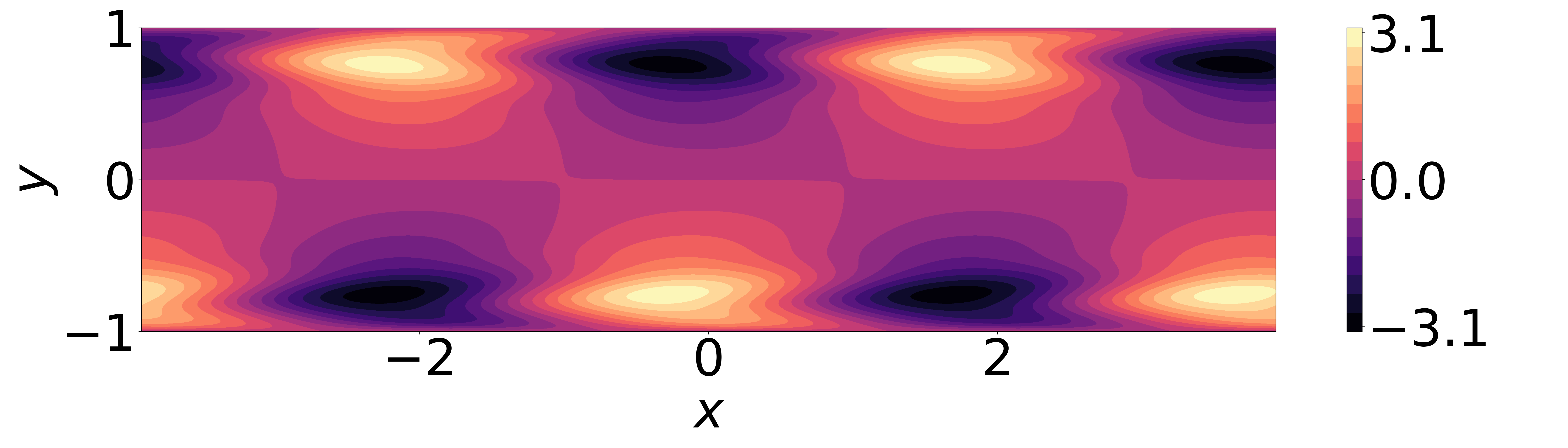}
         \caption{$\hat v_{x} (x,y)$}
         \label{fig:vx_f0d07_even_mode}
    \end{subfigure}
    \begin{subfigure}[b]{0.48\textwidth}
         \centering
         \includegraphics[width=\textwidth]{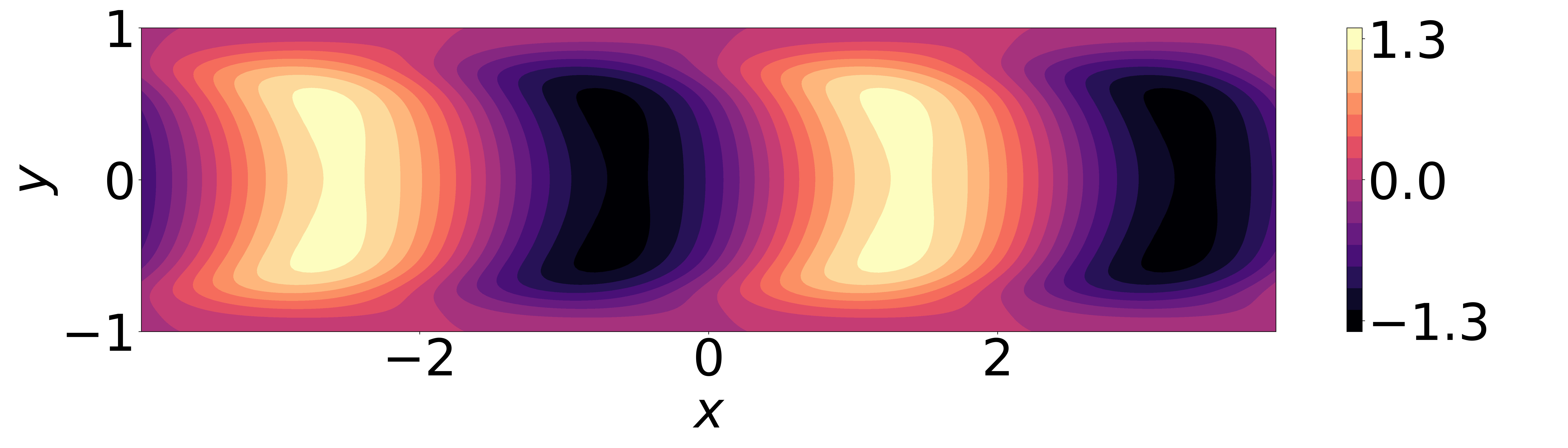}
         \caption{$\hat v_{y} (x,y)$}
         \label{fig:vy_f0d07_even_mode}
    \end{subfigure}
     \caption{Typical eigenfunctions of the shortwave mode shown at point `S' in figure \ref{example_three_modes}, where R=1000, $\alpha=1.6$.}  
 \label{ei2}
\end{figure}

\begin{figure}
    
     \centering
          \begin{subfigure}[b]{0.48\textwidth}
         \centering
         \includegraphics[width=\textwidth]{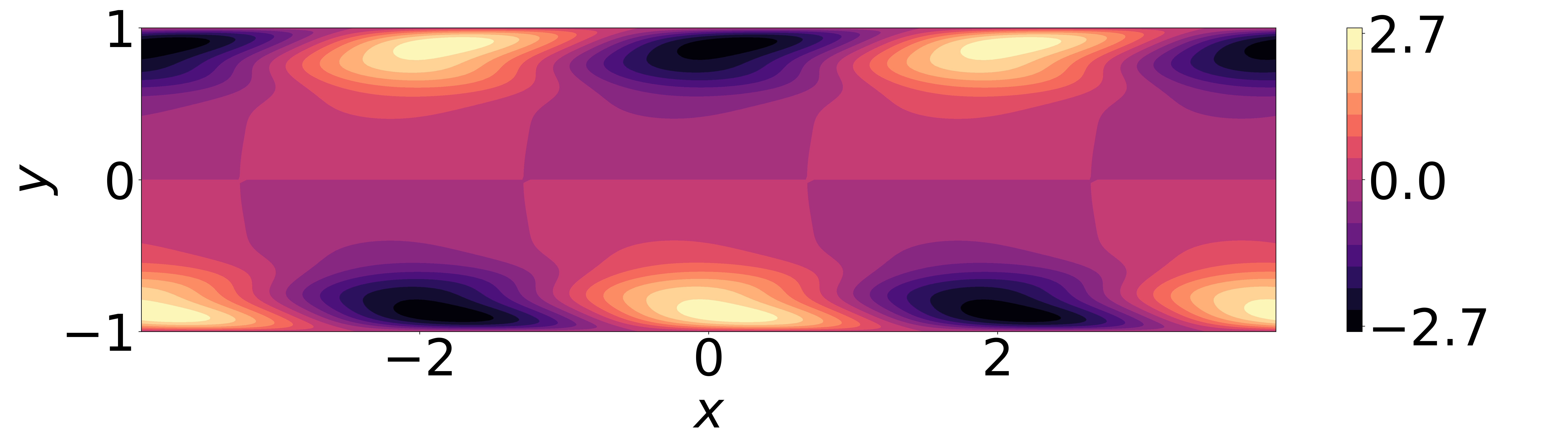}
         \caption{$\hat u_{x} (x,y)$}
         \label{fig:ux_R_1000_al_1d6_f0d07_even_mode_minimal}
     \end{subfigure}
     \begin{subfigure}[b]{0.48\textwidth}
         \centering
         \includegraphics[width=\textwidth]{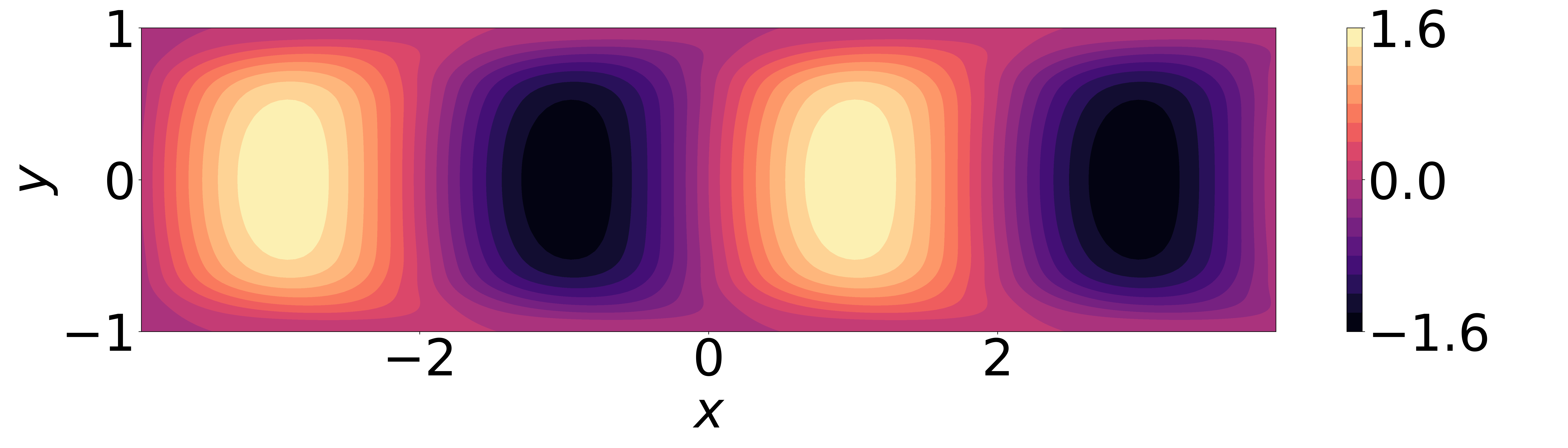}
         \caption{$\hat u_{y} (x,y)$}
         \label{fig:uy_R_1000_al_1d6_f0d07_even_mode_minimal}
     \end{subfigure}
    \hfill
       \begin{subfigure}[b]{0.48\textwidth}
         \centering
         \includegraphics[width=\textwidth]{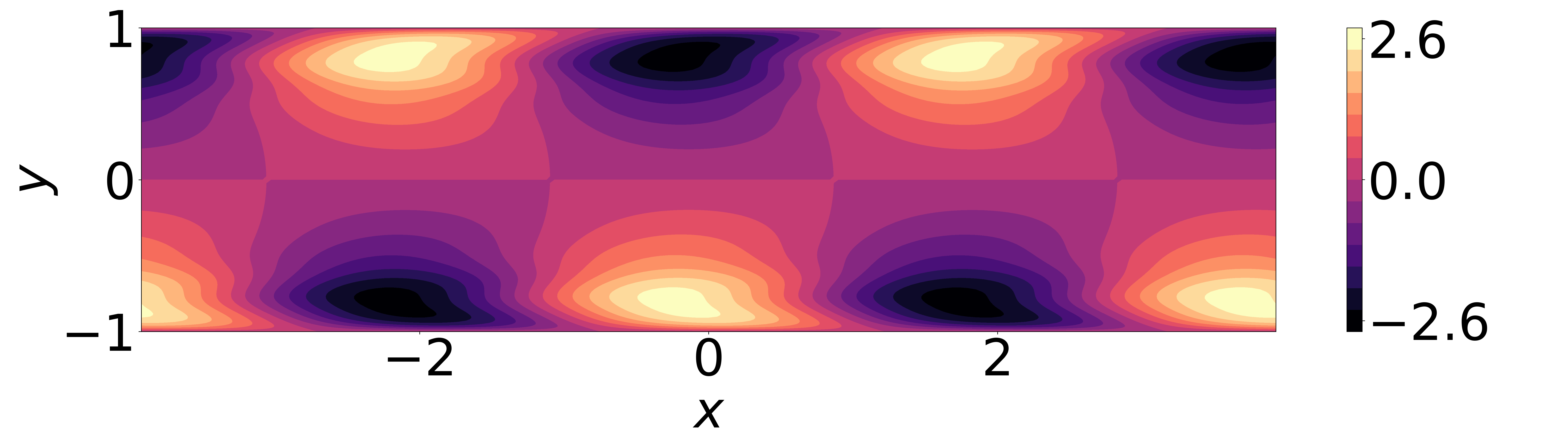}
         \caption{$\hat v_{x} (x,y)$}
         \label{fig:vx_R_1000_al_1d6_f0d07_even_mode_minimal}
    \end{subfigure}
    \begin{subfigure}[b]{0.48\textwidth}
         \centering
         \includegraphics[width=\textwidth]{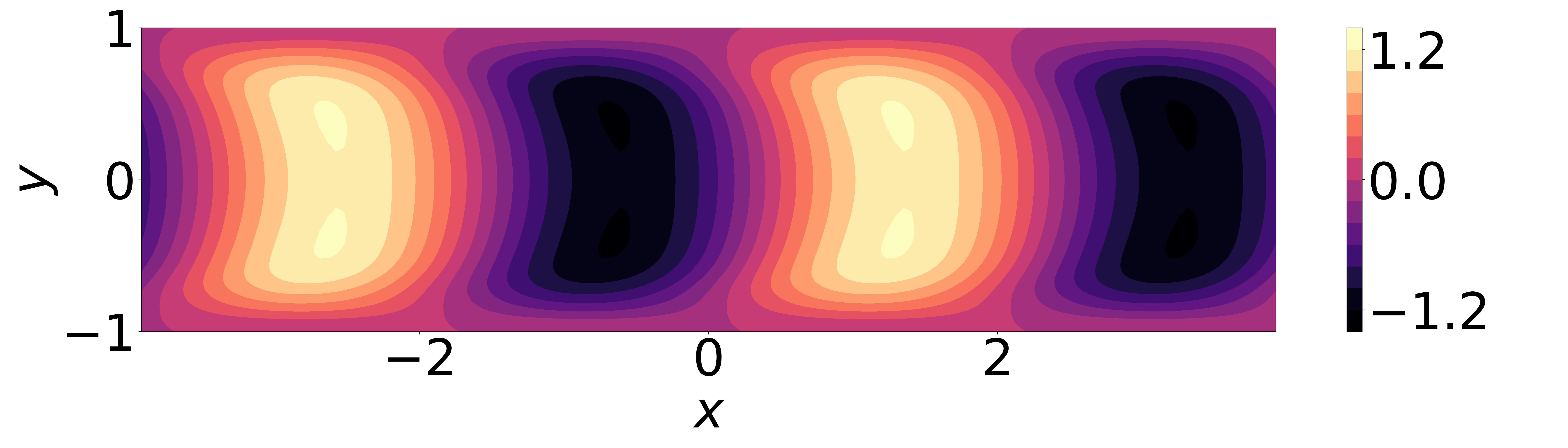}
         \caption{$\hat v_{y} (x,y)$}
         \label{fig:vy_R_1000_al_1d6_f0d07_even_mode_minimal}
    \end{subfigure}
     \caption{Characteristic eigenfunctions of the shortwave mode, obtained from the minimal composite equation (\ref{minimal_equation}) using identical parameters as those employed for Figure (\ref{ei2}).}  
 \label{ei4}
\end{figure}
\begin{figure}
     \centering
           \begin{subfigure}[b]{0.48\textwidth}
         \centering
         \includegraphics[width=\textwidth]{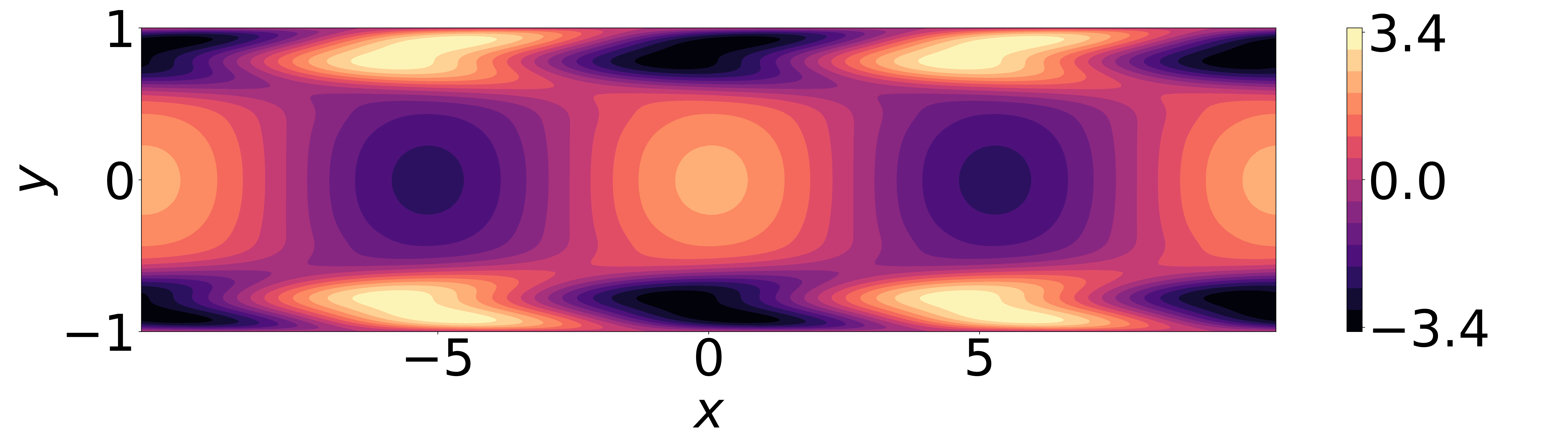}
         \caption{$\hat u_{x}(x,y)$}
         \label{fig:ux_f0d07_odd_mode}
     \end{subfigure}
     \begin{subfigure}[b]{0.48\textwidth}
         \centering
         \includegraphics[width=\textwidth]{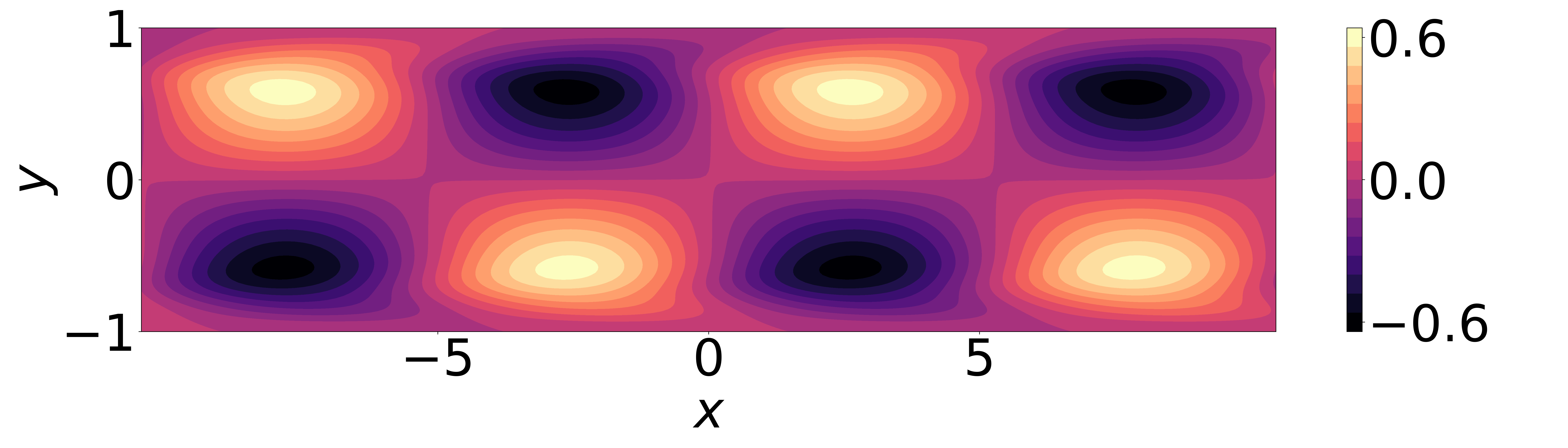}
         \caption{$\hat u_{y}(x,y)$}
         \label{fig:uy_f0d07_odd_mode}
     \end{subfigure}
    \hfill
    \begin{subfigure}[b]{0.48\textwidth}
         \centering
         \includegraphics[width=\textwidth]{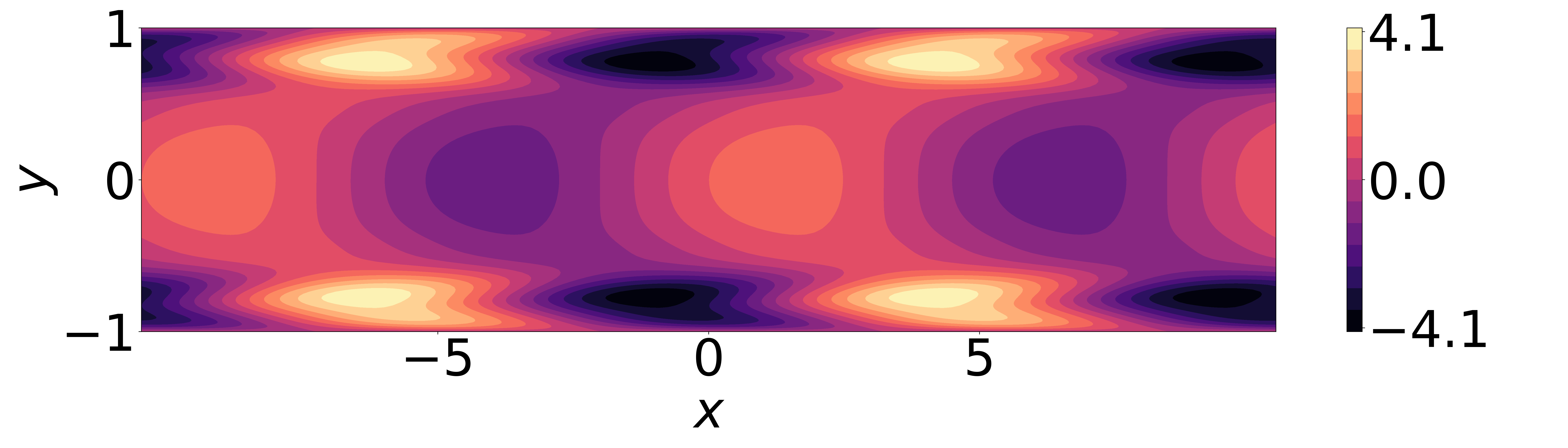}
         \caption{$\hat v_{x}(x,y)$}
         \label{fig:vx_f0d07_odd_mode}
    \end{subfigure}
    \begin{subfigure}[b]{0.48\textwidth}
         \centering
         \includegraphics[width=\textwidth]{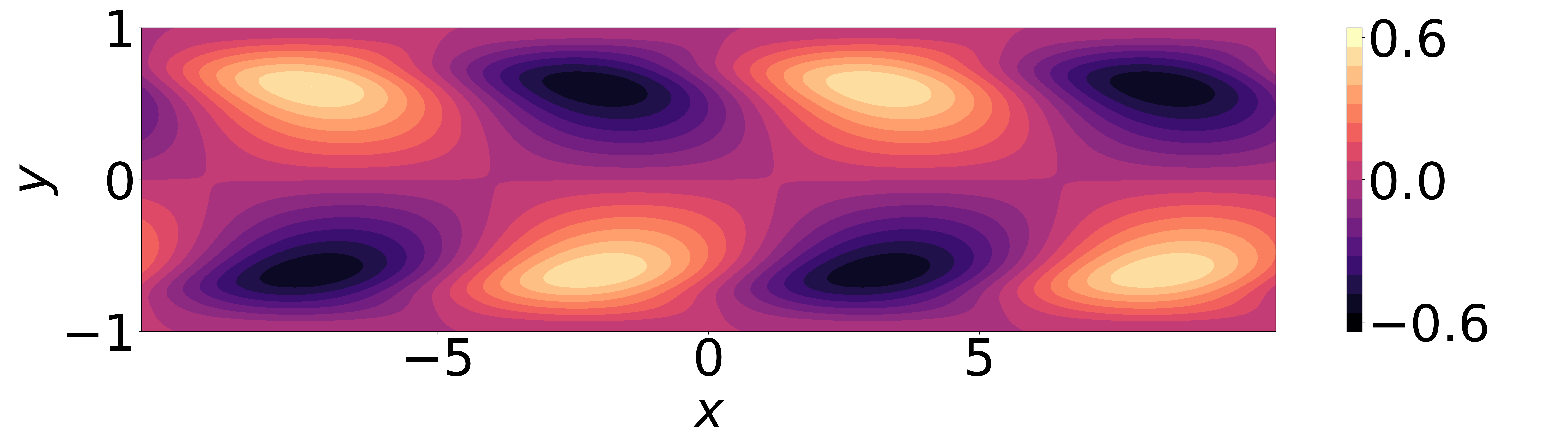}
         \caption{$\hat v_{y} (x,y)$}
         \label{fig:vy_f0d07_odd_mode}
    \end{subfigure}
     \caption{Typical eigenfunctions of the longwave mode, at point `L' in figure \ref{example_three_modes}, where R=3000, $\alpha=0.6$.}
 \label{ei3}
\end{figure}
\begin{figure}
     \centering
           \begin{subfigure}[b]{0.48\textwidth}
         \centering
         \includegraphics[width=\textwidth]{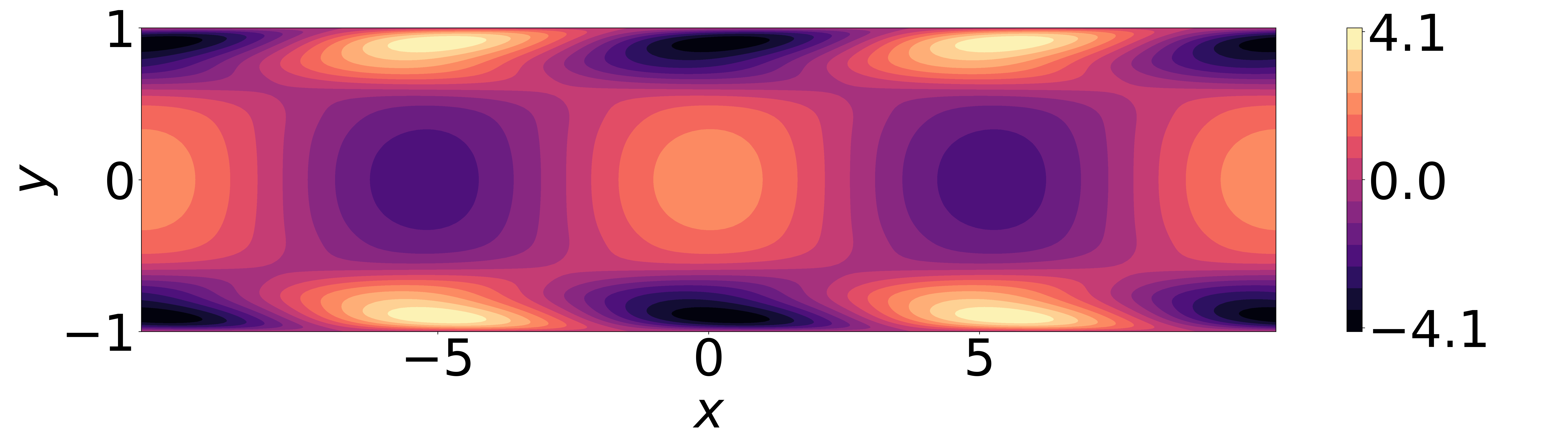}
         \caption{$\hat u_{x}(x,y)$}
         \label{fig:ux_R_3000_al_0d6_f0d07_odd_mode_minimal}
     \end{subfigure}
     \begin{subfigure}[b]{0.48\textwidth}
         \centering
         \includegraphics[width=\textwidth]{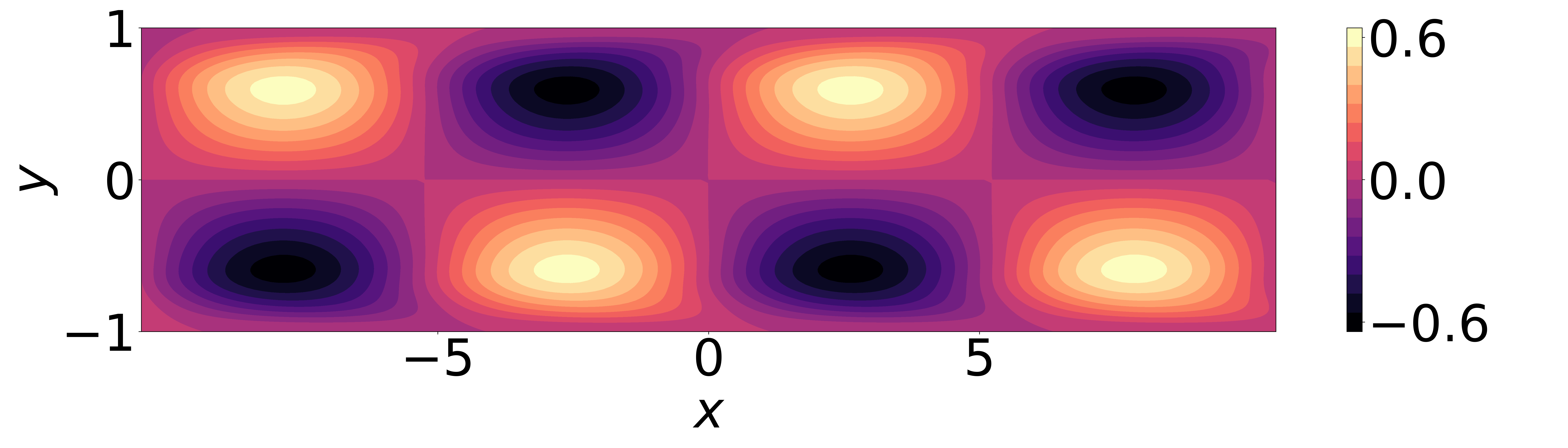}
         \caption{$\hat u_{y}(x,y)$}
         \label{fig:uy_R_3000_al_0d6_f0d07_odd_mode_minimal}
     \end{subfigure}
    \hfill
    \begin{subfigure}[b]{0.48\textwidth}
         \centering
         \includegraphics[width=\textwidth]{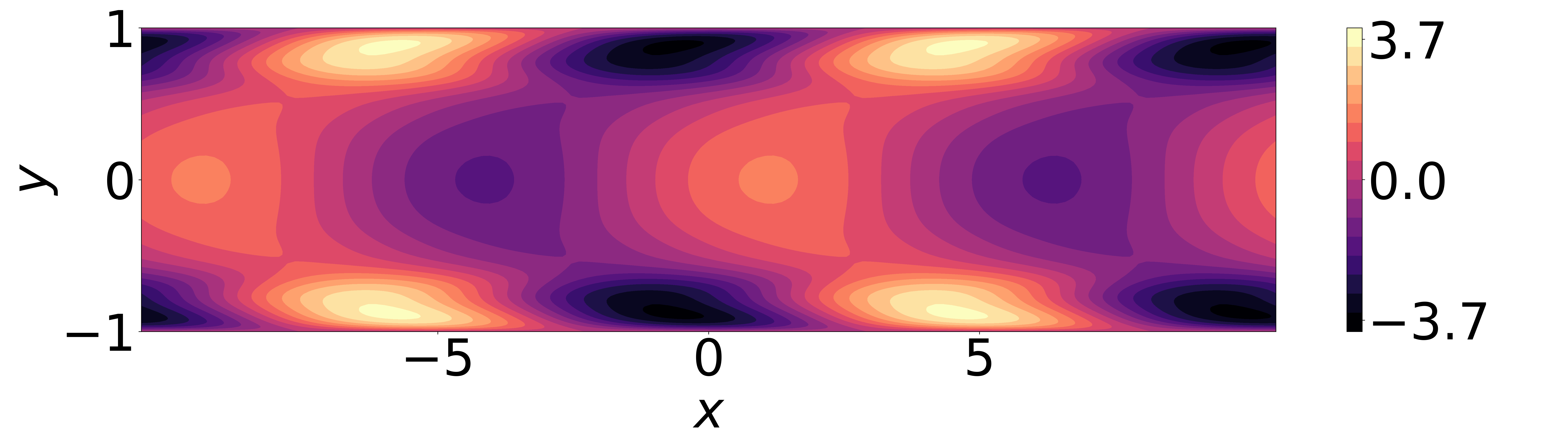}
         \caption{$\hat v_{x}(x,y)$}
         \label{fig:vx_R_3000_al_0d6_f0d07_odd_mode_minimal}
    \end{subfigure}
    \begin{subfigure}[b]{0.48\textwidth}
         \centering
         \includegraphics[width=\textwidth]{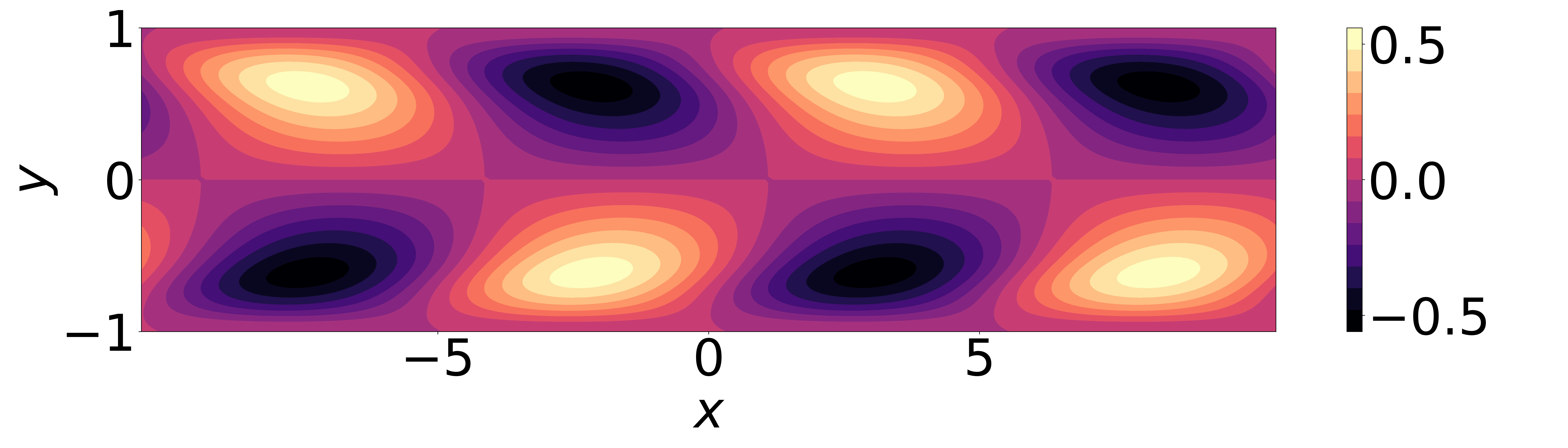}
         \caption{$\hat v_{y} (x,y)$}
         \label{fig:vy_R_3000_al_0d6_f0d07_odd_mode_minimal}
    \end{subfigure}
     \caption{Characteristic eigenfunctions of the longwave mode  obtained from the minimal composite equation (\ref{minimal_equation}) using identical parameters as those employed for Figure (\ref{ei3}).}
 \label{ei5}
 
\end{figure}
The very fact that the longwave mode is odd and the shortwave even means that their eigenfunctions are completely different in character, as seen in figures \ref{ei1} to \ref{ei5}. The eigenfunctions have been normalised to set the maximum value of the stream function $\psi$ to unity. The shortwave instability in figure \ref{ei2} displays much stronger streamwise velocity fluid perturbations than the TS. Also the near-wall structure of the streamwise velocity presents a distinct, wider and more symmetric reverse arrowhead shape than the TS. The eigenfunctions in the particle perturbation velocity components are strikingly different in the shortwave and the TS modes. In the shortwave mode, the particle dynamics is seen to follow the dynamics of the fluid, especially as seen in the streamwise velocity components. The normal velocity component is more peaky for the particles and more rounded for the flow. In contrast, the particles in the TS mode show a thinner region of strong streamwise velocity, but their wall-normal velocity is everywhere weak. The fact that the relevant portions of the eigenfunction profiles are thicker for the shortwave than for the TS mode is a consequence of the lower Reynolds numbers in the former, and we expect this from our critical-layer analysis above. For the shortwave mode, we compare the eigenfunctions from the full solution in figure \ref{ei2} to those from the minimal composite equation in figure \ref{ei4}. Though the arrowhead shape is now distorted, the overall similarity in the eigenfunction structure between the two is striking. This is strong visual evidence that the dominant physics is contained in the minimal composite theory.

The eigenstructure of the longwave instability is shown in figure \ref{ei3} from the full equation. Again the eigenfunctions from minimal composite theory, given in figure \ref{ei5}, are strikingly similar to those of the complete solution.
The energy budgets in the following section will be rendered very surprising, given how different the odd and even modes are in their eigenstructure.

     \section{Energy production and the critical layer}

\begin{figure}
     \centering
         \begin{subfigure}[b]{0.48\textwidth}
         \centering        \includegraphics[width=\textwidth]{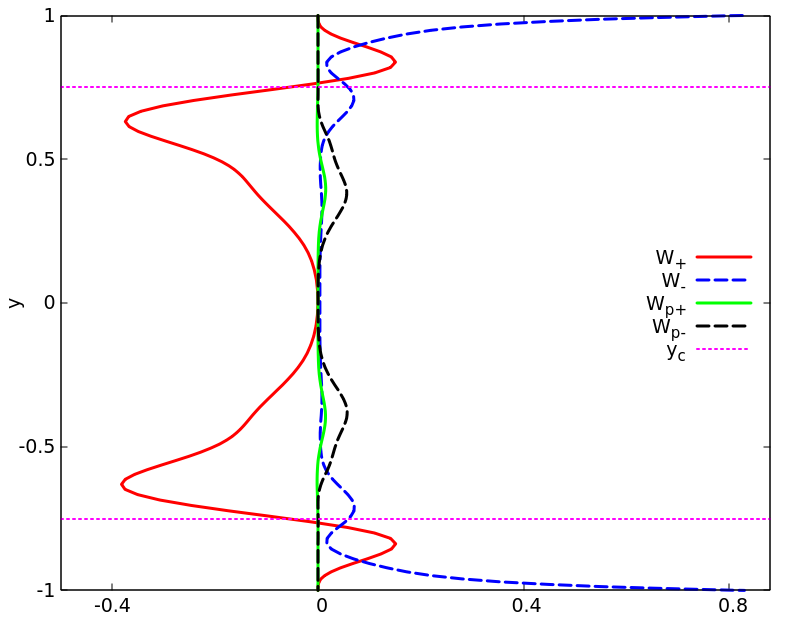}
         \caption{}
         \label{energy_al_1d56_R_1000_a0_0d40_new}
         \end{subfigure}
         \begin{subfigure}[b]{0.48\textwidth}
         \centering        \includegraphics[width=\textwidth]{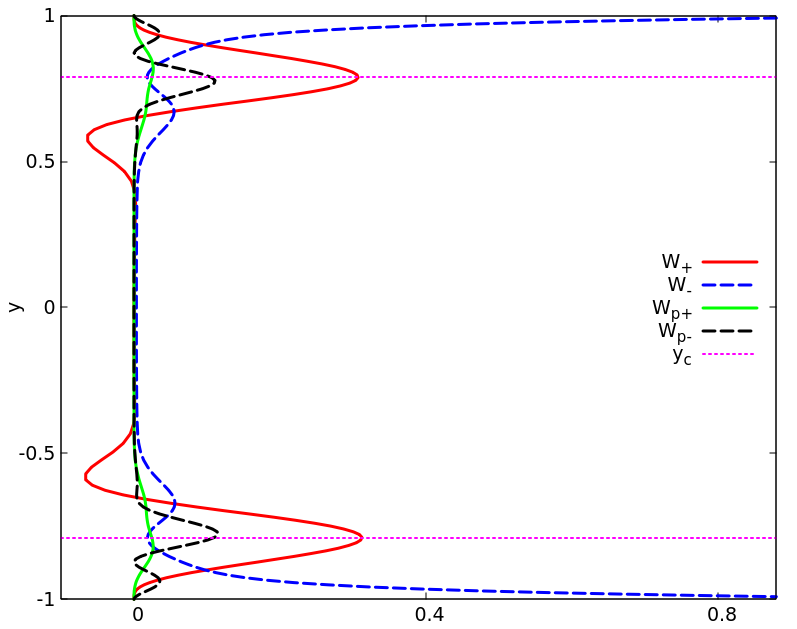}
         \caption{}
         \label{energy_al_1d56_R_1000_a0_0d75_new}
         \end{subfigure}
\caption{Contributions to the perturbation kinetic energy balance at a Reynolds number of $1000$ and a streamwise wavenumber of $\alpha=1.56$, with $f_{max}=0.30$, $\sigma=0.1$ and $S=2.5*10^{-4}$. The kinetic energy production $W_+$ due to the fluid is net negative in (a), where $a_p=0.40$, but net positive in (b) which is under overlap conditions, with $a_p=0.75$. The kinetic energy production has a noticeable contribution within the critical layer from particles, $W_{p+}$, in (b) but not in (a). The dissipation $W_-$ in the flow is similar in the two figures. The location $y=y_c$ is shown by the dashed pink lines.}  
\label{fig:energy}
\end{figure}
\begin{figure}
     \centering
         \begin{subfigure}[b]{0.48\textwidth}
         \centering        \includegraphics[width=\textwidth]{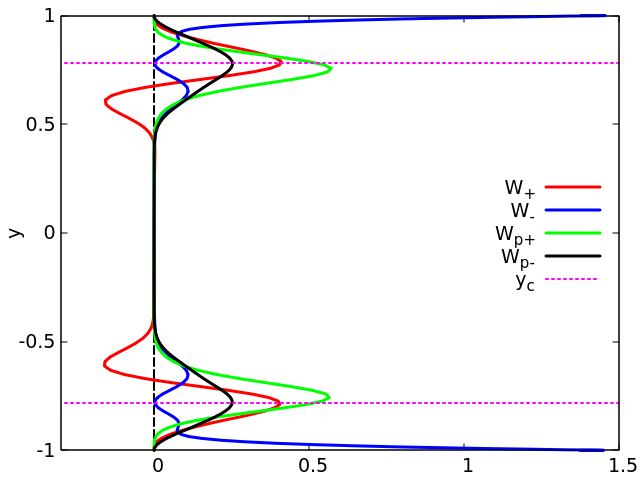}
         \caption{}
         \label{energy_al_0d6_R_3000_a0_0d75_new}
         \end{subfigure}
         \begin{subfigure}[b]{0.48\textwidth}
         \centering        \includegraphics[width=\textwidth]{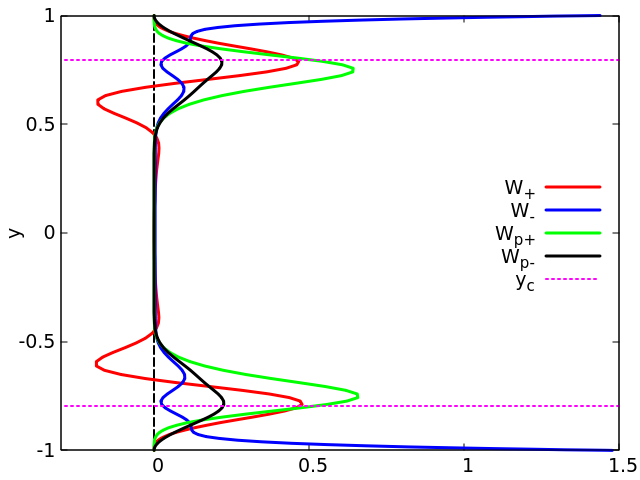}
         \caption{}
         \label{energy_al_1d6_R_1000_a0_0d75_new}
         \end{subfigure}
\caption{Comparison of contributions to the energy budget in the odd mode (a) at point `L' in figure \ref{example_three_modes}, where $R=3000$ and $\alpha=0.6$, with that of the even mode (b) at point `S' in figure \ref{example_three_modes}, i.e., $R=1000$ and $\alpha=1.6$. In both, $f_{\text{max}}=0.70$, $\sigma=0.1$, $S=8 \times 10^{-4}$, and $a_p=0.75$. Both of these modes are unstable, with net production beating dissipation by a small amount.}  
\label{fig:energy_odd_even}
\end{figure}

Figure \ref{fig:energy} shows the profiles across the channel of the four quantities that contribute to the growth of perturbation kinetic energy, written down in equation \ref{energy}. In the heavy particle limit, the two quantities $W_{\mu_{1}}$ and $W_{\mu_{2}}$ are zero. The parameters are all identical in figures \ref{fig:energy}(a) and \ref{fig:energy}(b), except for $a_p$, which corresponds to non-overlap conditions in (a) and overlap conditions in (b). The quantities plotted, when combined in the form given in equation \ref{energy}, and integrated over $y$ across the channel, in case (a) give a negative number, i.e., the perturbation is highly damped, whereas this results in a positive number in case (b), indicating an exponentially growing mode. The dissipation quantities $W_-$ and $W_{p-}$ are positive definite by definition. The striking difference between the two figures is in the production of perturbation kinetic energy by the fluid ($W_+$). In the non-overlap case, the net production is clearly negative, i.e., $W_+$ is feeding back kinetic energy from the perturbations to the mean flow, and contributing to the decay of the perturbations. Under overlap conditions on the other hand, the production is sharply peaked and positive in the critical layer, leading to the instability. Thus we establish that moving the the particle-laden layer from non-overlap to overlap conditions has a remarkable effect on the solutions to the linear stability problem. Also noticeable is that under non-overlap conditions, there is effectively no contribution to energy production from the particles, i.e., $W_{p+}$ is too small to matter. But in the overlap case, particles contribute directly to the instability as well as by triggering the fluid production. In both cases, $W_{p-}$ is small, and concentrated in the particle-laden layer, while in both cases the fluid dissipates perturbation kinetic energy primarily near the walls (see $W_-$), i.e., displays classical behaviour. 

Figure \ref{fig:energy_odd_even}, comparing the energy budgets of the odd and even mode of overlap instability, holds a surprise. Note that by equation (\ref{energy}) this plot is constructed entirely from the eigenfunctions depicted in figures \ref{ei2} and \ref{ei3}. The eigenfunctions are completely different in structure, but the energy budgets are very close to each other. Closer observation reveals that the two eigenfunctions are indeed very similar in the neighborhood of the critical and wall layers, which explains the similarity in the production and dissipation. This finding begs the question of the possibility of different eigenstructures in the bulk in different flows yielding critical-layer driven instabilities. We are not aware of any other such situation in shear flows. 

\section{Viscosity stratification}
\begin{figure}
  \centerline{\includegraphics[width=0.95\textwidth]{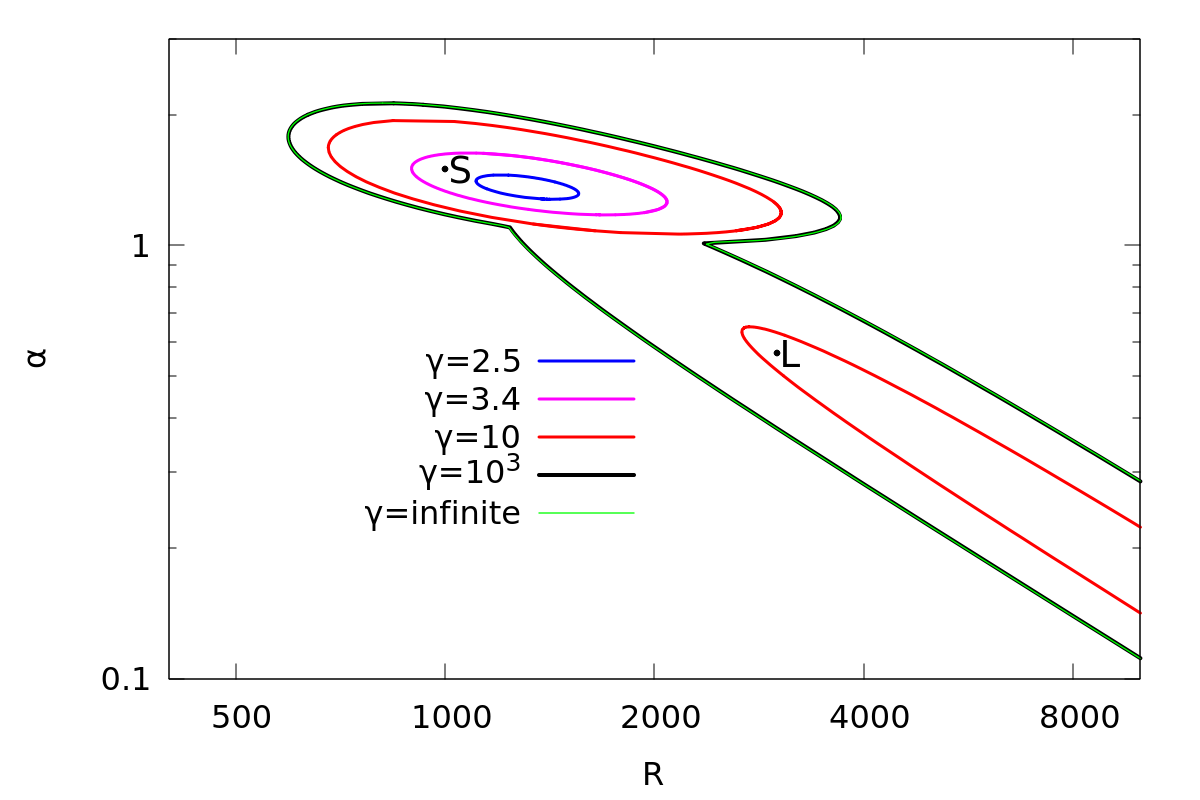}}
    \caption{Neutral boundaries of the shortwave and longwave modes of instability for various density ratios $\rho_{p}/\rho_{f}=2.5\gamma$. All other parameters are as in figure (\ref{example_three_modes}), where we had $\gamma \to \infty$.  }
    \label{fig:neutral_visc_strat}
\end{figure}
Thus far we have worked in the heavy particle limit, where the mass fraction is finite and the volume fraction of the particles is negligible. We now relax this, and impose a finite particle to fluid density ratio, thereby allowing viscosity to vary in accordance with equation (\ref{einstein}). The stability equation (\ref{eq:visc_strat}) is now applicable, and the mean flow profile $U(y)$ is given by:
\begin{equation} 
    (\bar{\mu}U^{\prime})^{\prime\prime}=0,
\end{equation} with the boundary conditions $U(\pm1)=0$, and $U(0)=1$.
The effect on the neutral boundaries of the viscosity variation is seen in Fig. \ref{fig:neutral_visc_strat} to be uniformly stabilizing in this flow, with both stability boundaries shrinking significantly as $\gamma$ is increased. The longwave mode vanishes below $\gamma=10$, while a small region of instability persists in the shortwave mode up to $\gamma=2.5$. The maximum particle volume fraction we have considered occurs for this $\gamma$, which is 11 percent at the maximum in the particle layer and lower elsewhere.
\begin{figure}
    \begin{subfigure}{0.48\textwidth}
          {\includegraphics[width=\textwidth]{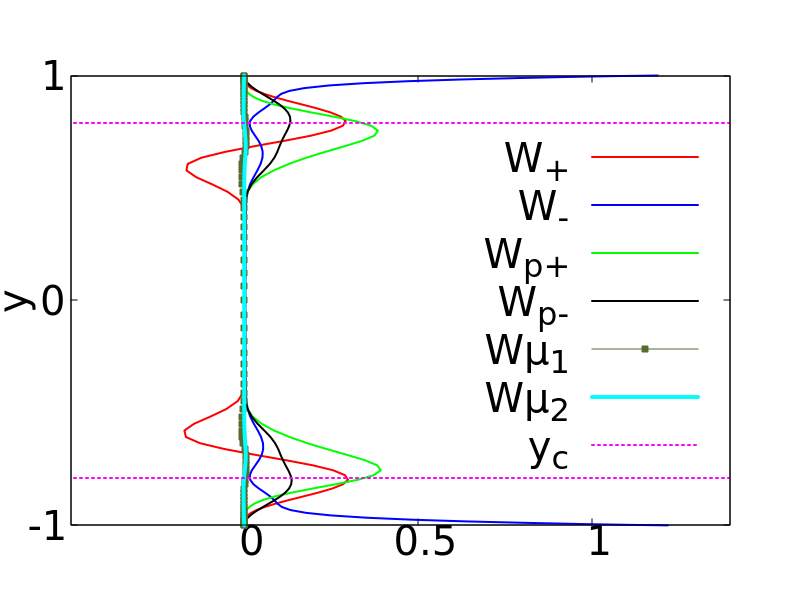}} 
          \caption{}
    \end{subfigure}
    \begin{subfigure}{0.48\textwidth}
        {\includegraphics[width=\textwidth]{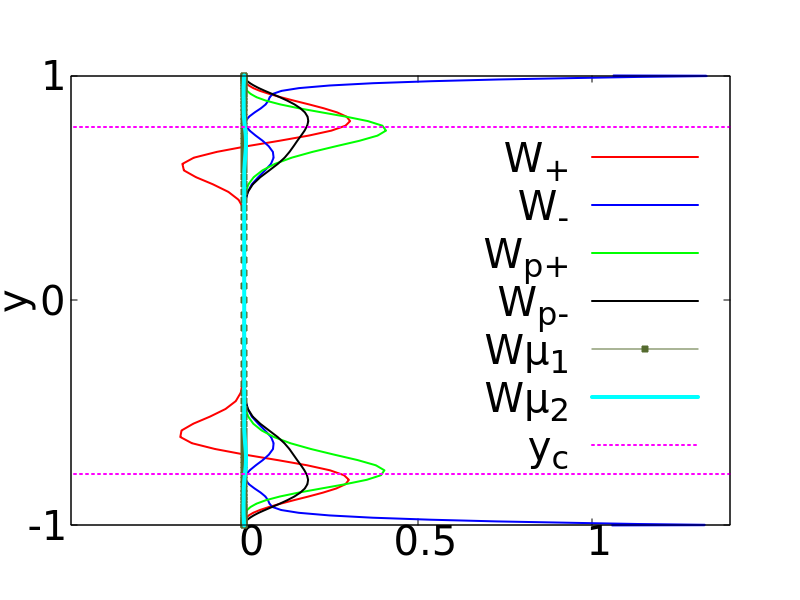}}
        \caption{}
    \end{subfigure}

    \caption{Energy budget for $\gamma=3.4$. Profiles of the quantities in equation \ref{energy}  are shown. (a) R=1000, $\alpha=1.6$, see point `S' in Fig. \ref{fig:neutral_visc_strat}, and (b) R=3000, $\alpha=0.6$, i.e., point `L' in that figure. At this $\gamma$, `S' is still slightly unstable, whereas `L' is completely stabilized.}
    \label{fig:Energy_ffac_3d4}
\end{figure}
We ask why this large change happens, since in the critical layer analysis told us that viscosity stratification should not enter the dominant balance at these modest volume fractions. The energy budget for the two modes S and L, which are now stable, is shown for $\gamma=3.4$ in Fig. \ref{fig:Energy_ffac_3d4}. We note that the production is lower than seen for no viscosity stratification in Fig. \ref{fig:energy_odd_even}. But there s practically no contribution from the viscosity variation terms $W_{\mu_{1}}$ and $W_{\mu_{2}}$. The change in stability is entirely due to the change in the mean profile $U$.

\section{Summary and outlook}

For decades, shear flows have thrown up surprises in their stability behaviour, and the different mechanisms of instability, though not easy to predict, are crucial to unravel. This is an important reason why these flows are appealing to study. We have persevered to show that the inclusion of particles in a Poiseuille flow is such a case, where we present the mechanism of low Reynolds number instability.

We have shown that the response of the flow to non-uniform particle loading may be divided into two broad categories that we term overlap and non-overlap conditions. Under non-overlap conditions, the particle-laden layer lies at some distance from the critical layer, where perturbation kinetic energy is produced, and particles do not significantly alter this process. However, when there is an overlap between these layers, there is a dramatic alteration of stability behaviour, with two modes of instability apart from the TS mode appearing. The fundamental difference between overlap and non-overlap conditions is starkly visible in Fig. \ref{general} and has been discussed above. Though these modes have been observed in one older study \cite{rudyak1997hydrodynamic} at constant viscosity, they had not been explained before, to our knowledge. The shortwave overlap mode occurs at much lower Reynolds number than the TS mode, and supports wavelengths of the order of the channel width. The longwave overlap mode appears over a wide range of Reynolds numbers and supports wavelengths which could be as small as the channel width but become longer and longer with increasing Reynolds number. This mode is rather unusual in that it is odd in the wall-normal component of the perturbation velocity. The three modes of instability show regimes of distinct existence, and go through interesting intersections and mergers with changes in parameters. 

We derive the lowest-order critical and wall layer equations for particulate parallel shear flow for dilute particle loading, and show how they differ from the classical equations for clean flow. This is combined with an energy-budget analysis which brings out the consequences for stability. The reason for the existence of two categories of behaviour is shown to lie in the dynamics within the critical layer. Variations in the base particle concentration within the critical layer significantly alter the production of disturbance kinetic energy. The result is a large destabilisation for this loading profile under a range of conditions. The wall layer is seen not to be a major player. To directly evaluate the lowest-order physics, we derive a minimal composite equation, which contains all the terms in the complete stability equations which contribute at the leading order somewhere in the flow, i.e., in the outer, critical or wall layers. The wall layer contributes no additional terms not present in the other two. The minimal composite equation is shown to contain the essential physics of the overlap instabilities, in terms of trends in the critical Reynolds number and indeed in the eigenfunction behaviour.

In the limit of heavy particles, the volume loading is negligible, so the viscosity is constant. We then consider finite particle to fluid density ratios, where the volume loading is finite but small. Now viscosity varies with particle concentration. The change in the mean flow velocity profile effects a significant stabilisation, whereas the explicit viscosity gradient terms are shown to be non-players in this case. Whether this is a consequence of the special viscosity profile that our loading produces remains to be studied in the future. This question is interesting because in the case of viscosity variations produced by temperature or solute concentration, an overlap mode of instability was predicted by \cite{ranganathan2001stabilization,govindarajan2004effect} and seen in experiments such as those of \cite{hu2018viscous}. Related overlap physics can change the nature of turbulence and the transition to turbulence in heated flow \citep{zonta2024turbulence}. 

The important next question therefore is whether the location of particle loading can affect the transition to turbulence in shear flows. Non-modal linear effects might be important for certain ranges of parameters in this process, and need further attention, in the overlap and non-overlap regimes. Interestingly, when the Reynolds number is about two thousand, nonhomogeneous loading shows exponential growth whereas homogeneous loading \citep{klinkenberg2014linear} shows merely transient growth, indicating that the route to turbulence in the two can be different. Direct numerical simulations are needed to determine the route to turbulence and the possibility of multiple routes due to the different modes of instability. Finally, any theoretical treatment of particulate flow is almost always rife with assumptions whose validity needs to be established by detailed experiments. The effects of geometry are not obvious either, and need investigation. Behavior opposite to present findings is seen in pipe flow experiments by \citet{Guazzelli2003_p1,Guazzelli2003_p2}. It is worth noting that pipe flow differs from channel flow in many aspects. Notably, Squire’s theorem, which we have shown here to hold true for channel flow, is not applicable there, so helical modes are often the least stable.

\medskip

\appendix
\section{Derivation of linear stability equations}

Upon linearization of equations (\ref{NS_eq}-\ref{particle_continuity_eq})  we have
\begin{equation}
\begin{aligned}
  \left(\frac{\partial \mathbf{\hat{u}}}{\partial t}+ \mathbf{\hat{u}} \cdot \mathbf{\nabla}\mathbf{U}+\mathbf{U} \cdot \mathbf{\nabla}\mathbf{\hat{u}}\right)&=-\mathbf{\nabla} \hat{p}+\frac{1}{R}\bigg[\mathbf{\nabla}\bar{\mu}\cdot\left(\mathbf{\nabla}\mathbf{\hat{u}}+\left( \mathbf{\nabla}\mathbf{\hat{u}} \right)^T\right)+ 
\mathbf{\nabla}\hat{\mu}\cdot\left(\mathbf{\nabla}\mathbf{U}+\left( \mathbf{\nabla}\mathbf{U} \right)^T\right)\bigg]\\
&+\frac{\bar{f}}{SR}\left(\mathbf{\hat{v}}-\mathbf{\hat{u}}\right),  
\end{aligned}
    \label{NS_eq_linear}
\end{equation}
\begin{equation}
     \mathbf{\nabla}\cdot\mathbf{\hat{u}}=0,
     \label{Continuity_eq_linear}
\end{equation}
\begin{equation}
    \left(\frac{\partial \mathbf{\hat{v}}}{\partial t}+ \mathbf{\hat{v}} \cdot \mathbf{\nabla}\mathbf{U}+\mathbf{U} \cdot \mathbf{\nabla}\mathbf{\hat{v}}\right)=-\frac{1}{SR}\left(\mathbf{\hat{v}}-\mathbf{\hat{u}}\right),  
    \label{particle_eq_linear}
\end{equation}
\begin{equation}
\frac{\partial \hat{f}}{\partial t}+\mathbf{\nabla}\cdot (\bar{f} \, \mathbf{\hat{v}})+\mathbf{\nabla}\cdot (\hat{f} \, \mathbf{U})=0.
\label{particle_continuity_eq_linear}
\end{equation}

We start by performing a normal mode analysis, considering single Fourier modes for the perturbation quantities ($\mathbf{\hat{u}}$, $\mathbf{\hat{v}}$,  $\hat{p}$, $\hat{f}$, $\hat{\mu}$) in both the $x$-direction and $z$-direction, as well as in time. In other words, these quantities are represented as ($\hat{u}$, $\mathbf{\hat{v}}$,  $\hat{p}$, $\hat{f}$, $\hat{\mu}$) = $(\mathbf{u}, \mathbf{v},p,f,\mu) \exp\left[i(\alpha(x-ct)+\beta z)\right]$. 

We may now express the particle velocity in terms of the flow velocity using equations (\ref{particle_eq_linear}), to get
\begin{equation}
      (v_{x},v_{y},v_{z})=({\cal M}u_{x}-SR{\cal M}^{2}U^{\prime}u_{y},{\cal M}u_{y},{\cal M}u_{z})
      \label{v_in_terms_u}
\end{equation}
where 
\begin{equation}
    {\cal M}=\frac{1}{1+i\alpha SR(U-c)}.
\end{equation} 
Additionally, by taking the divergence of equation (\ref{NS_eq_linear}), we write the pressure laplacian in terms of the velocity field as
\begin{equation}
    \begin{aligned}
    -\nabla^2 p&=2i\alpha U^{\prime}u_{y}-\bigg[ 2\bar{\mu}^{\prime}\nabla^{2}u_{y}+2\bar{\mu}^{\prime\prime}Du_{y}+2i\alpha U^{\prime}D\mu+2i\alpha U^{\prime\prime}\mu\bigg]\\
    &-\frac{1}{SR}\bigg[\bar{f^{\prime}}({\cal M}-1)u_{y}+\bar{f}{\cal M}^{\prime}u_{y}-i\alpha SR{\cal M}^{2}U^{\prime}\bar{f}u_{y}\bigg].
    \end{aligned}
    \label{pressure_eq}
\end{equation}
We apply the operator $\nabla^2$ to the $y$ component of the vector equation (\ref{NS_eq_linear}), use equations (\ref{v_in_terms_u}) and (\ref{pressure_eq}), and divide throughout by $-i\alpha^2$ to express the resulting equation in terms of the variables $u_y$ and $\mu$: 
\begin{equation}
\begin{aligned}
     -ic \nabla^{2}\dfrac{u_{y}}{-i\alpha } &=i\left[U^{\prime\prime} -U\nabla^{2}\right]\dfrac{u_{y}}{-i\alpha }+\frac{1}{\alpha R}\left[\bar{\mu}'' \left( -\nabla^2 + 2 D^2 \right) + 2 \bar{\mu}' D \nabla^2 + \bar{\mu} \nabla^4 \right] \dfrac{u_{y}}{-i\alpha } \\
     &+\frac{1}{\alpha R} \left[U^{\prime\prime\prime} + 2 U^{\prime\prime} D - U^{\prime} \nabla^2 + 2 U^{\prime} D^2 \right] \mu \\
     & +i\bigg[({\cal M}^{2}U^{\prime}\bar{f})^{\prime}-(U-c){\cal M}\bar{f}^{\prime}D-(U-c){\cal M}\bar{f}\nabla^{2} \bigg]\dfrac{u_{y}}{-i\alpha}
\end{aligned}
\label{uy_mu_eq_f}
\end{equation}
 If we express the particle velocity in terms of the flow velocity using the equation (\ref{v_in_terms_u}) and subsequently utilizing the continuity equation (\ref{Continuity_eq_linear}), we can convert equation (\ref{particle_continuity_eq}) into an expression that includes the variables $u_{y}$ and $\mu$, leading, after dividing throughout by $-i\alpha$, to 
\begin{equation}
    (U-c)\gamma\mu+\bigg[- RS{\cal M}^{2}U^{\prime}\bar{f}^{\prime}+ \frac{({\cal M}\bar{f})^{\prime}}{i\alpha}\bigg] u_{y}=0.
    \label{mu_eq}
\end{equation}

Equations (\ref{uy_mu_eq_f}) and (\ref{mu_eq}) represent the linear stability equations for three-dimensional  perturbations. If we substitute $(\alpha^2+\beta^2)=\alpha_{2D}^2$, $\alpha R=\alpha_{2D} R_{2D}$, and $u_{y}/\alpha =u_{y,2D}/\alpha_{2D}$ into the aforementioned equations (\ref{uy_mu_eq_f}) and (\ref{mu_eq}), these equations become equivalent to those of a two-dimensional system with the wave number denoted as $\alpha_{2D}$, the Reynolds number as $R_{2D}$ and the velocity eigenfunction $u_{y,2D}$. We thus show that Squire's theorem may be extended to dusty channels with inhomogeneous loading including viscosity variation as well.
Thus for two-dimensional perturbations, these equations become (\ref{eq:visc_strat}) and (\ref{eq:visc_part_conti}).

\medskip

\section{Dominant balance in the wall layer }
In shear flows, the critical and wall layers are often not well-separated, and the overlap mode of instability presents such a case. We conduct the exercise below only to confirm that wall effects are not bringing in new physics into the instability. As before we define inner variables 
\begin{equation}
    \xi=\frac{y-y_w}{\epsilon_w},\quad {\rm and} \quad \lambda=\frac{y-y_w}{\delta_w}
    \end{equation}
where $y_w=\pm1$ are the wall locations. Also, we have $U_w\equiv U(y_w)=0$ and $U_{w}^{'}\equiv U^{\prime}(y_w)$.
We expand the variables in the form
\begin{equation} 
 u_y=\sum_{n=0}^{\infty} \epsilon_{w}^n  u_{y, n}(\xi), \quad v_y  =\sum_{n=0}^{\infty} \delta_{w}^n v_{y,n}(\lambda) \quad {\rm and} \ \  v_x  =\sum_{n=0}^{\infty} \delta_{w}^n v_{x, n}(\lambda).
\label{expan_wall}
\end{equation}
At the lowest order, $u_{y,0}=0$, and $v_{y,0}$ is proportional to this quantity and thus vanishes.
At the next order, after some algebra, we obtain the scaling $\delta_w=\epsilon_w$, and we can use the incompressibility condition $D_\xi  u_{y, 1} =- i \alpha u_{x,0}$. Additionally, we have the following equations:
\begin{equation}
    v_{x,0}=\dfrac{u_{x,0}}{1-i\alpha cSR},
\end{equation}
\begin{equation}
        v_{x,1}=\frac{u_{x,1}}{1-i\alpha cSR}-\left(\frac{SRU_{w}^{\prime}}{1-i\alpha cSR}\right)v_{y,1}, \quad  {\rm and} \quad v_{y,1}=\frac{u_{y,1}}{1-i\alpha cSR}.
    \label{wall_layer_vxy}
\end{equation}
We can express the dominant-balance composite equation for the flow as follows:
\begin{equation}
    \left[-c\bigg(1+ \frac{\bar{f}}{1-i\alpha cSR}\bigg) D_{\xi}^{2} + i\frac{1}{\alpha R\epsilon_w^2}D_{\xi}^4 - \frac{\epsilon_w}{\sigma}\bigg(\frac{cD_{\chi}\bar{f}}{1-i\alpha cSR}\bigg)D_{\xi}\right]u_{y,1} = 0
    \label{wall_layer_uy_eq}
\end{equation}
The structure of equation (\ref{wall_layer_uy_eq}) in the wall layer is the same as (\ref{composite}) in the critical layer. There are changes in the coefficients, which change the scaling of $\epsilon_w$ to be $(\alpha R)^{-1/2}$. Again, when $\epsilon_w \sim \sigma$, along with significant overlap of the wall layer and the particle layer, the first derivative of the particle concentration profile is among the biggest terms at the lowest order. In our range of study, the wall layer is always very thin and well-separated from the critical layer. All three terms are important when $\epsilon_w \sim (\alpha R)^{-1/2}$. In our investigation, we always positioned the particle layer at a considerable distance from the wall layer. Consequently, the mass fraction $\bar{f}$ and its derivative become negligible in the wall layer, and the dominant balance (\ref{wall_layer_uy_eq}) is unaffected by the presence of particles.

\begin{acknowledgments}
{\bf Acknowledgments:} Our research at ICTS-TIFR is supported by the Department of Atomic Energy, Government of India, under Project No. RTI4001.
\end{acknowledgments}

\bibliographystyle{jfm}
\bibliography{jfm-instructions}
\end{document}